\documentclass[doublespace]{elsart}
\usepackage{graphicx}
\usepackage{subfigure}
\usepackage{amsmath, amssymb}
\usepackage{epsf}
\usepackage{pstricks}
\usepackage{ccaption}

\makeatletter\@addtoreset{equation}{section}\makeatother
\def\theequation{\thesection.\arabic{equation}}
                                                                                
\newcommand \be{\begin{eqnarray}}
\newcommand \ee{\end{eqnarray}}
\newcommand \ba{\begin{eqnarray}}
\newcommand \ea{\end{eqnarray}}
\newcommand \bs{\boldsymbol}
\newcommand \mc{\mathcal}

\newcommand \E{{\bs{\mc E}}}
\newcommand \mE{{\mc E}}

\captionnamefont{\small\sf}
\captiontitlefont{\small\sf}

\begin{document}
\small

\begin{frontmatter}

\title{An Adaptive Grid Refinement Strategy for the Simulation 
of Negative Streamers}
\author[cwi]{C. Montijn\thanksref{nwo}}
\ead{carolynne.montijn@cwi.nl}
\thanks[nwo]{Financial support is provided by NWO in the
Computational Science program.} 
\author[cwi]{W. Hundsdorfer\thanksref{bricks}}
\ead{willem.hundsdorfer@cwi.nl}
\thanks[bricks]{This research was also supported by the Dutch 
national program BSIK: knowledge and research capacity, in the
ICT project BRICKS, theme MSV1.}
\author[cwi,tue]{U. Ebert\thanksref{bricks}}
\ead{ute.ebert@cwi.nl}
\address[cwi]{CWI, P.O. Box 94079, 1090 GB Amsterdam, The Netherlands}
\address[tue]{Department of Physics, TU Eindhoven, 5600 MB Eindhoven, 
The Netherlands}

\begin{abstract}

The evolution of negative streamers during electric breakdown of 
a non-attaching gas can be described
by a two-fluid model for electrons and positive ions. It consists of
continuity equations for the charged particles including drift, 
diffusion and reaction in the local electric field, coupled to
the Poisson equation for the electric potential.  
The model generates field enhancement and steep propagating 
ionization fronts at the tip of growing ionized filaments.
An adaptive grid refinement method for the simulation of these structures  
is presented. It uses finite volume  spatial discretizations
and explicit time stepping, which allows the decoupling of 
the grids for the continuity equations from those for the Poisson equation. 
Standard refinement methods in which the refinement criterion is based 
on local error monitors fail due to the pulled character of the streamer 
front that propagates into a linearly unstable state. We present 
a refinement method which deals with all these features.
Tests on one-dimensional streamer fronts as well as on three-dimensional 
streamers with cylindrical symmetry (hence effectively 2D for numerical purposes)
are carried out successfully. Results on fine grids are presented, they show 
that such an adaptive grid method is needed to capture the streamer 
characteristics well.
This refinement strategy enables us to adequately compute
negative streamers in pure gases in the parameter regime
where a physical instability appears: branching streamers.
\end{abstract}


\end{frontmatter}

\section{Introduction}

When non-ionized or lowly ionized matter is exposed to high electric
fields, non-equilibrium ionization processes, so-called discharges, occur.
They may appear in various forms depending on the spatio-temporal 
characteristics of the electric field and on the pressure and volume 
of the medium. One distinguishes the dark,
glow or arc discharges that are stationary, and transient
non-stationary phenomena such as leaders and streamers.
We will focus here on streamers, that are growing filaments of plasma
whose dynamics are controlled by highly
localized and nonlinear space charge regions.  

Streamers occur in nature as well as in many technical applications.
They play a role in creating the path of sparks and lightning~\cite{baz2000}
and they are believed to be directly observable as the multiple channels
in so-called sprite discharges. These huge, lightning related
discharges above thunderclouds attract large research effort
since their first observation in 1989~\cite{fra1990,sen1995,ger2000,pas2002-2}. 
Because of the reactive radicals they emit, streamers are used for the treatment of 
contaminated media like exhaust gasses~\cite{cle1989,vel2000}, polluted 
water~\cite{sat1996,abo2002} or biogas~\cite{nai2004}. More recently, 
efforts have been undertaken to improve the flow around aircraft wings by coupling
space charge regions to gas convection~\cite{boe2005}. 

Streamers can be either cathode directed or anode directed; the charge
in their head is then positive or negative, respectively.
Positive streamers propagate against the drift velocity 
of electrons, therefore they need additional (and not well known)
mechanisms like nonlocal photoionization or background ionization.
Negative streamers in simple non-attaching gases like nitrogen or argon, 
on the other hand, can be described by a minimal model with rather 
well-accessible parameters~\cite{ebe1997}. We therefore focus on this case.

The minimal streamer model for negative streamers is a continuum approximation
for the densities of the electrons and positive ions with a local
field-dependent impact ionization term and with particle diffusion and 
particle drift in the local electrical field. As space charges of the
streamer change the field, and the field determines drift and reaction
rates of the particles, the model is nonlinear, and steep ionization fronts
between ionized and non-ionized regions emerge dynamically.

After the first claim \cite{arr2002} that a streamer filament 
within this deterministic continuum model in a sufficiently high field 
can evolve into an unstable state where it branches spontaneously,
streamer branching was observed in more 
simulations~\cite{roc2002,liu2004,liu2006}.
While the physical nature of the Laplacian instability was elaborated 
in simplified analytical models~\cite{arr2002,arr2004,meu2004,meu2005,ebe2005}, 
other authors challenged the accuracy of the numerical 
results~\cite{kul2000,pan2001,kul2002,pan2003}. 
Based on purely numerical evidence, their questions were justified, 
as all simulations within the minimal model up to now were carried 
out on uniform grids, and the numerical convergence on finer grids 
could hardly be tested.

Therefore, in the present paper, we present a grid refinement strategy
for the minimal streamer model and discuss its results.
This procedure allows us to test the numerical convergence of branching,
and also to calculate streamers efficiently in longer systems without 
introducing too many grid points.
Moreover, the simulations in~\cite{arr2002,roc2002} were 
performed on the same uniform grid for both the particle densities
and the electric potential. As the Poisson equation for the electric
potential has to be solved on the complete large non-ionized outer space, 
this grid had to be much larger than the actual domain in which 
the particles evolve. Furthermore, the streamer has an inner layered
structure with very steep ionization fronts and thin space charge
layers. An efficient refinement is therefore badly needed. 

There is an additional complication: standard grid refinement 
procedures in regions with steep gradients fail due to the pulled 
character of the streamer front.
``Pulling'' means that the dynamics and, in particular,
the front velocity, is determined in the linearly unstable high field 
region ahead of the front rather than by the regions with the steepest
density gradients~\cite{ebe1997,ebe2000}. The high field region
where any single electron immediately will create an ionization avalanche,
is generated by the approaching curved ionization front itself.

In the paper, a refinement strategy dealing efficiently with all
these specific problems is developed. It is based on physical
insight on the one hand, by which the relevant region for the particle
densities can be restricted, and on the knowledge of the importance of
the leading edge on the other hand. The drift-diffusion equations for the
particle densities and the Poisson equation for the electric potential
are treated separately, allowing the solutions for particle densities
and electric potential to adapt to the specific difficulties.  
This refinement procedure will be applied to problems in two spatial 
dimensions, or in three dimensions with cylindrical symmetry.
                                                                   
The outline of this paper is as follows. First, in Section~2, we give a
brief description of the model. In Section~3 the numerical discretizations
are introduced and motivated. Section~4 discusses the refinement procedure.
Section~5 contains a one-dimensional example in which some of the
essential numerical difficulties with local grid refinements of pulled 
fronts are illustrated and discussed.  Section~6 deals with 
the performance of our refinement algorithm,
and the results are presented in the Section~7.
The final section contains a discussion of the results and conclusions.
\section{Hydrodynamic approximation for the ionized channel}
\subsection{Drift, diffusion and ionization in a gas}
                                                                                                                              
The essential properties of anode directed streamer propagation in a non-attaching gas
like $N_2$ or $Ar$, can be analyzed by a fluid model for two species of charged particles
(electrons and positive ions). It consists of continuity equations for the electron and positive
ion number densities, $n_e$ and $n_+$,
\ba
\displaystyle\frac{\partial n_e}{\partial t}&-\nabla\cdot(n_e\mu_e{\bf E}+D_e\nabla n_e) & =  S_i \, ,\label{neeq}  \\
\displaystyle\frac{\partial n_+}{\partial t}& & =  S_i \, . \label{n+eq}
\ea
The electrons drift with a velocity $\mu_e{\bf E}$ and diffuse with a diffusion coefficient $D_e$. 
Here ${\bf E}$ is the local electric field and $\mu_e$ the electron mobility. The ions can be considered to 
be immobile on the short time scales considered here, since their mobility is two orders of magnitude 
smaller than that of the electrons~\cite{dha1987}.  We remark that extending our algorithm 
to moving ions and eventually other species is rather straightforward, and not including
this makes no difference for the algorithm itself. 

The source term $S_i$ represents creation of electrons and positive ions through impact 
ionization, and is the same for both species since charge is conserved during an 
ionization event. The impact ionization can be described with Townsend's approximation~\cite{rai1991}
(but any other local field dependence can be inserted as well)
\be
S_i=n_e\mu_e|{\bf E}|\alpha(|{\bf E}|)=n_e\mu_e|{\bf E}|\alpha_0\,e^{-E_0/|{\bf E}|}\, ,
\label{towneq}
\ee
where $\alpha_0$ is the ionization coefficient and $E_0$ the threshold field for ionization.
We do not include photoionization or recombination since, in the
particular case of $N_2$ and for the short time scales considered, these processes are negligible compared
to impact ionization~\cite{pan2001}. However, an ionization mechanism such as
photoionization is essential for the development of positive streamers,
which are therefore excluded from the present study.

The electric field ${\bf E}$ is determined through Poisson's equation for
the electric potential $V$,
\be
\nabla^2V=\frac{e}{\epsilon_0}(n_e-n_+)\, , \,\,\,\,\,\,{\bf E}=-\nabla V \, ,
\label{poisseq}
\ee
where $\epsilon_0$ is the permittivity of free space, and $e$ the electron charge.
                                                                                                                              
\subsection{Dimensional analysis}
This model has been implemented in dimensionless form. The characteristic length
and field scales emerge directly from Townsend's ionization formula~(\ref{towneq})
as $l_0=\alpha_0^{-1}$ and $E_0$, respectively. The characteristic velocity is then given
as $v_0=\mu_eE_0$, which leads to a characteristic timescale $t_0=l_0/v_0=l_0/(\mu_e E_0)$.
The characteristic diffusion coefficient then becomes $D_0=l_0^2/t_0$. The number density scale
emerges from the Poisson equation~(\ref{poisseq}), $n_0=\epsilon_0 E_0/el_0$.
We use the values from~\cite{dha1987,vit1994} for $\mu_e$, $E_0$ and $\alpha_0$ in $N_2$ at 300 K,
which depend on the neutral gas density,
\be
\mu_e      \simeq  \frac{380}{(N/N_0)}\,\,\,  \frac{{\rm cm^2}}{{\rm V\,s}}\, ,\,\,
\alpha_0   \simeq  \frac{4332}{(N/N_0)}\,\,\, \frac{1}{\rm cm}\, , \,\,
E_0        \simeq  \frac{2\cdot 10^5}{(N/N_0)}\,\,\,  \frac{\rm V}{\rm cm}\, .
\ee
Inserting these values in the characteristic scales we obtain, for molecular nitrogen at 
normal conditions,
\be
\displaystyle l_0\simeq\frac{2.3\cdot 10^{-4}}{(N/N_0)}\,\,\,{\rm cm}\, ,\,\,\,\,\,\,
\displaystyle t_0\simeq\frac{3\cdot 10^{-12}}{(N/N_0)}\,\,\,{\rm s}\, ,\,\,\,\,\,\,
\displaystyle n_0\simeq\frac{4.7\cdot 10^{14}}{e\, ((N/N_0)^2}\,\,\,\frac{1}{\rm cm^3}\, ,\,\,\,\,\,\,
\displaystyle D_0\simeq\frac{1.8\cdot 10^4}{(T/T_0)}\,\,\, \frac{\rm cm^2}{\rm s}\, .
\ee
Here $N_0$ and $T_0$ are the neutral gas density and temperature under normal conditions.
The dimensionless quantities are then defined as follows,
\be
\displaystyle {\bf r}=\frac{{\bf R}}{l_0}\, ,\,\,\,\,\,\,  \displaystyle \tau=\frac{t}{t_0}\, ,\,\,\,\,\,\,  
  \displaystyle \sigma=\frac{n_e}{n_0}\, ,\,\,\,\,\,\, \displaystyle \rho=\frac{n_+}{n_0}\, ,\,\,\,\,\,\, 
\displaystyle {\bs{\mc E}}=\frac{{\bf E}}{E_0}\, ,\,\,\,\,\,\,
  \displaystyle\phi=\frac{V}{(E_0l_0)}\, ,\,\,\,\,\,\,\displaystyle D=\frac{D_e}{D_0}\, .
\ee
For the diffusion coefficient we use the value given in~\cite{dut1975}, $D_e$=1800 cm$^2$ s$^{-1}$,
which gives a dimensionless diffusion coefficient of 0.1 under normal conditions.

Inserting these dimensionless quantities into the continuity equations~(\ref{neeq})-(\ref{poisseq}),
we obtain
\ba 
\displaystyle\frac{\partial \sigma}{\partial \tau} & = & \nabla\cdot(\sigma{\bs{\mathcal E}}+D\nabla\sigma) 
                + \sigma|{\bs{\mathcal E}}|\exp(-1/|{\bs{\mathcal E}}|) \, , \label{sigmaeq}\\[4mm]
\displaystyle\frac{\partial\rho}{\partial \tau} & = & \sigma|{\bs{\mathcal E}}|\exp(-1/|{\bs{\mathcal E}}|) \, ,\label{rhoeq}\\[4mm]
\displaystyle\nabla^2\phi& = &\sigma-\rho\, , \quad {\bs{\mathcal E}}=-\nabla\phi\, ,\label{phieq}
\ea
where $\sigma$ and $\rho$ denote the dimensionless electron and ion number densities, respectively, $\tau$
the dimensionless time,
${\bs{\mathcal E}}$ the dimensionless electric field, $\phi$ the dimensionless electric potential 
and $D$ the dimensionless diffusion coefficient.

We refer to the equations~(\ref{sigmaeq})-(\ref{phieq}) as 
the minimal streamer model, since it contains all the basic physics needed
for negative streamers in a non-attaching gas.

\subsection{Geometry and boundary conditions}
\label{subsecbc}
In narrow geometries, streamers frequently are growing from pointed
electrodes, that create strong local fields in their neighborhood
and a pronounced asymmetry between the initiation of positive and negative
streamers~\cite{vel2002}. On the other hand, in many natural discharges
and, in particular, for sprites above thunderclouds \cite{liu2004},
it is appropriate
to assume that the electric field is homogeneous. Of course, dust
particles or other nucleation centers can play an additional role in discharge generation,
but in this paper we will focus
on the effect of a homogeneous field on a homogeneous gas.

The computational domain is limited by two planar electrodes. The model is implemented
in a cylindrical symmetric coordinate system $(r,z)\in(0,L_r)\times(0,L_z)$, such 
that the electrodes are placed perpendicular to the axis of symmetry ($r$=0), the 
cathode at $z=0$ and the anode at $z=L_z$.
The boundary conditions for the electric potential read
\be
\phi(r,0,\tau) = 0 \,, \qquad
\phi(r,L_z,\tau) = \phi_0>0 \,, \qquad
\frac{\partial\phi}{\partial r}(L_r,z,\tau) = 0 \,.
\label{bcphi}
\ee
The background electric field then becomes
\be
{\E_b}=-|\E_b|{\bf \hat{e}_z}=-\frac{\phi_0}{L_z}{\bf \hat{e}_z}\, ,
\ee
where ${\bf\hat{e}_z}$ is the unit vector in the $z-$direction.
The radial boundary at $L_r$ is virtual, and only present to create a finite computational domain. In
order for the boundary condition no to affect the solution near the axis of symmetry along which the 
streamer propagates, we need to place this boundary relatively far from the axis of symmetry.

Throughout this article, we use a Gaussian initial ionization seed, placed on 
the axis of symmetry, at a distance $z=z_b$ from the cathode,
\be
\sigma(r,z,0)=\rho(r,z,0)=\sigma_0\exp\left(\frac{r^2+(z-z_b)^2}{R_b^2}\right)\, .
\label{ic}
\ee
The maximal density $\sigma_0$ of this seed, the radius $R_b$ at which the 
density drops to 1/e of its maximal value, and the value of $z_b$
differ from case to case, and will be specified where needed. 
Furthermore, the use of 
a dense seed, in particular in low fields, accelerates the emergence of a streamer.
We remark that the initial seed is charge neutral. 

The continuity equation for the electron density is second order in space, and 
therefore requires two boundary conditions for each direction in space.
At $r=L_r$ and $z=L_z$ we use Neumann boundary conditions,
so that electrons that arrive at those boundaries may flow out of or into the system,
but in all cases discussed in this paper the streamer is too far from the boundary for this to happen.
At the cathode, we impose either homogeneous Neumann or Dirichlet boundary conditions. In the first case
we again allow for a net flux of particles through the boundary. Dirichlet conditions will only be used for a one-dimensional
test in this paper.
To recapitulate, the boundary conditions for the
electrons read
\be
\frac{\partial\sigma}{\partial z}(r,0,\tau) = 0\, , \text{ or } \sigma(r,0,\tau) = 0\, ,\quad
\frac{\partial\sigma}{\partial z}(r,L_z,\tau) = 0\, ,\quad
\frac{\partial\sigma}{\partial r}(L_r,z,\tau) = 0\, .
\label{bcsigma}
\ee
We notice that,
if $z_b\gg R_b$, the ionization seed is detached from the cathode, and 
this results in practice in a zero inflow of electrons. On the other hand, 
placing the seed near the cathode will result in an inflow of electrons. Varying 
the value of $z_b$ will therefore enable us to investigate the influence 
of the inflow of electrons on the streamer propagation. 
\section{Numerical discretizations}
\label{secnumdisc}
                                                                                   
In our numerical simulations we shall mainly consider the streamer
model with radial symmetry, making it effectively two dimensional.
To illustrate some of the difficulties and their solutions we will also deal 
with the one-dimensional case.
                                                                                   
In the cylindrically symmetric coordinate system introduced in the previous section,
the equations~(\ref{sigmaeq})-(\ref{phieq}) read 
\ba
\frac{\partial \sigma}{\partial \tau}
&=& \displaystyle\frac{1}{r}\frac{\partial (r\sigma \mE_r)}{\partial r}
    + \frac{\partial(\sigma \mE_z)}{\partial z}
    + \frac{D}{r}\frac{\partial}{\partial r}(r\frac{\partial\sigma}{\partial r})
    + D \frac{\partial^2\sigma}{\partial z^2}
    + \sigma|{\E}|e^{-1/|{\E}|} \, , \label{cdrs2d}\\[4mm]
\displaystyle\frac{\partial\rho}{\partial \tau} &=&\displaystyle \sigma|{\E}|e^{-1/|{\E}|} \, , \label{cdrr2d}\\[4mm]
\displaystyle \nabla^2\phi & = & \displaystyle \frac{1}{r}\frac{\partial}{\partial r}
(r\frac{\partial \phi}{\partial r}) + \frac{\partial^2\phi}{\partial z^2}
\;=\; \sigma-\rho \, .
\label{poiss2d}
\ea
The electric field ${\E} = (\mE_r,\mE_z)^{\text T}$ can be computed
from the electric potential as
\be
{\E}  =  -\nabla\phi = -\Big(\frac{\partial\phi}{\partial r}, \,
\frac{\partial\phi}{\partial z} \Big)^{\rm T} \, .
\label{compe}
\ee
The boundary conditions for this system have been treated in Sect.\,\ref{subsecbc}.
                                                                                   
\subsection{Spatial discretization of the continuity equations}
\label{subsecdiscrcont}                                                                                   
The equations will be solved on a sequence of (locally) uniform
grids with cells
\be
{\mathcal C}_{ij} = [(i-1)\Delta r, i \Delta r] \times [(j-1)\Delta z, j \Delta z],\, i=1...M_r,\, j=1...M_z\, , 
\nonumber
\ee
where $M_r=L_r/\Delta r$ and $M_z=L_z/\Delta z$ are the number of grid points in
the $r$- respectively $z$-direction, 
and cell centers $(r_i,z_j) = ((i-\frac{1}{2})\Delta r,(j-\frac{1}{2})\Delta z)$.
We denote by $\sigma_{i,j}$ and $\rho_{i,j}$ the density approximations
in the cell centers.  These can also be viewed as cell averages.
The electric potential $\phi_{ij}$ and field strength
$|{\E}|_{i,j}$ are taken in the cell centers, whereas
the electric field components are taken on the cell vertices.
For the moment it is supposed that the electric field is known,
its computation will be discussed later on.
                                                                                   
The equations for the particle densities are discretized with finite
volume methods, based on mass balances for all cells.
Rewriting the continuity equations~(\ref{cdrs2d})-(\ref{cdrr2d})
will result in the semi-discrete system
\be
\begin{array}{rcl}
\displaystyle\frac{d\sigma_{i,j}}{d \tau} & = &
\displaystyle \frac{1}{r_i \Delta r}
(r_{i-\frac{1}{2}} F_{i-\frac{1}{2},j} - r_{i+\frac{1}{2}} F_{i+\frac{1}{2},j})
+ \frac{1}{\Delta z}
(F_{i,j-\frac{1}{2}}-F_{i,j+\frac{1}{2}})
+ S_{i,j} \, ,
\\[4mm]
\displaystyle\frac{d\rho{_{i,j}}}{d \tau} & = &\displaystyle S_{i,j} \, ,
\end{array}
\label{sds}
\ee
in which $F = F^a + F^d$. $F^a$ and $F^d$ are the advective and diffusive electron fluxes 
through the cell boundaries, and $S_{ij}$ is the source
term in the grid cell~${\mathcal C}_{ij}$.

The discretization of the advective terms requires care.
A first order upwind scheme as used in~\cite{pan2001,pan2003} appears to
be much too diffusive~\cite{bob1998}, leading to a totally different
asymptotic behavior on realistic grids. Moreover, it is expected that the
numerical diffusion might over-stabilize the numerical scheme, thereby
suppressing interesting features of the solutions. This explains why
streamer branching is not seen by the authors of~\cite{pan2001,pan2003}.
On the other hand, higher order linear discretizations lead to
numerical oscillations and negative values for the electron density,
that will grow in time due to the reaction term. This holds in
particular for central discretizations \cite{hun2003}.
The choice was made to use an upwind-biased scheme with flux limiting.
This gives mass conservation and monotone solutions without introducing
too much numerical diffusion.
For the limiter we will take the Koren limiter function, which is
slightly more accurate than standard choices such as the van Leer
limiter function~\cite{hun2003}.

Denoting $v^+ = \max(v,0)$ and $v^- = \min(v,0)$ to distinguish upwind
directions for the components of the drift velocity $v_r=-\mE_r$ and $v_z=-\mE_z$, the advective fluxes 
in the $r$- direction are computed by
\be
F^{a}_{i+\frac{1}{2},j}  = 
v^{+}_{r;\,i+\frac{1}{2},j} \Big[ \sigma_{i,j} + \psi\big(\theta_{i,j}\big)
\big(\sigma_{i+1,j}-\sigma_{i,j}\big)\Big]
+ \, v^{-}_{r;\,i+\frac{1}{2},j} \Big[ \sigma_{i+1,j}
+ \psi\big(\frac{1}{\theta_{i+1,j}}\big)\big(\sigma_{i,j}-\sigma_{i+1,j}\big)\Big],
\label{fluxa}
\ee
in which
\be
\theta_{i,j} = 
\frac{\sigma_{i,j}-\sigma_{i-1,j}}{\sigma_{i+1,j}-\sigma_{i,j}} \,, 
\qquad 
\psi(\theta) =
\max\Big(0,\min\Big(1,\frac{1}{3}+\frac{\theta}{6},\theta\Big)\Big) \, .
\ee
The advective fluxes in the vertical direction are computed in the same way;
the fact that the $r$-direction is radial is already taken care of in
(\ref{sds}).
Note that in regions where the solution is smooth we will have values
of $\theta_{ij}$ close to~$1$, and then the scheme simply reduces
to the third-order upwind-biased discretization corresponding to
$\psi(\theta) = \frac{1}{3} + \frac{1}{6}\theta$. In non-smooth regions where
monotonicity is important the scheme can switch to first-order upwind,
which corresponds to $\psi(\theta) = 0$.

The diffusive fluxes are calculated with standard second-order central 
differences as
\be
F^d_{i+\frac{1}{2},j} = \frac{D}{\Delta r}(\sigma_{i,j}-\sigma_{i+1,j}) \, .
\qquad
F^d_{i,j+\frac{1}{2}} = \frac{D}{\Delta z}(\sigma_{i,j}-\sigma_{i,j+1}) \, ,
\ee
Finally, the reaction term $S_{ij}$ in (\ref{sds}) is
computed in the cell centers as
\be
S_{i,j} = \sigma_{i,j}|{\E}|_{ij} e^{-1/|{\E}|_{ij}} \,.
\ee

Boundary values will be either homogeneous Dirichlet or homogeneous Neumann type. 
For example, for Dirichlet boundary conditions $\sigma=0$ for $z=0$, we introduce virtual values~\cite{hun2003}
\be
\sigma_{i,0}=-\sigma_{i,1}\, ,\quad \sigma_{i,-1}=-\sigma_{i,2}\, .
\ee
For Neumann boundary conditions $\partial_z\sigma=0$ for $z=0$ we set
\be
\sigma_{i,0}=\sigma_{i,1}\, ,\quad \sigma_{i,-1}=\sigma_{i,2}\, .
\ee
These formulas follow from the approximations 
\be
\begin{array}{rcl}
\sigma(r_i,z_{\frac{1}{2}}) & = & \displaystyle\frac{1}{2}(\sigma(r_i,z_0)+\sigma(r_i,z_1))+\mc O(\Delta z^2)\\
\sigma_z(r_i,z_{\frac{1}{2}}) & = & \displaystyle\frac{1}{\Delta z}(\sigma(r_i,z_1)-\sigma(r_i,z_0))+\mc O(\Delta z^2)
\end{array}
\ee

\subsection{Spatial discretization of the Poisson equation}
                                                                                           
The electric potential $\phi$ is computed through a second-order central
approximation of Eq.\,(\ref{phieq}), and is defined at the cell
centers:
\be
\sigma_{i,j}-\rho_{i,j} = 
  \displaystyle\frac{\phi_{i+1,j}-2\phi_{i,j}+\phi_{i-1,j}}{\Delta r^2}+
               \frac{\phi_{i+1,j}-\phi_{i-1,j}}{2r_i\Delta r}
 &\,\,\, + &\displaystyle \frac{\phi_{i,j+1}-2\phi_{i,j}+\phi_{i,j-1}}{\Delta z^2}\,.
\ee

The electric field components are then computed by using a second-order
central discretization of ${\E}=-\nabla\phi$, they are defined in the cell
boundaries,
\be
\mE_{r;i+\frac{1}{2},j} = \frac{\phi_{i,j}-\phi_{i+1,j}}{\Delta r}\, , \qquad
\mE_{z;i,j+\frac{1}{2}} = \frac{\phi_{i,j}-\phi_{i,j+1}}{\Delta z}\, .
\label{erzdisc}
\ee

The electric field strength is computed at the cell centers, therefore the 
components are first determined in the cell centers by averaging the cell 
boundary values, and the electric field strength then becomes:
\be
|{\E}|_{i,j} = 
\frac{1}{2}\sqrt{(\mE_{r;i-\frac{1}{2},j}+\mE_{r;i+\frac{1}{2},j})^2 +
(\mE_{z;i,j-\frac{1}{2}}+\mE_{z;i,j+\frac{1}{2}})^2}\, .
\label{edisc}
\ee

We notice here that, discretizing $\nabla\cdot{\E}$ with a second-order 
central scheme gives
\be
\frac{\partial}{\partial \tau}(\nabla\cdot{\E})  =
                   \frac{\partial(\sigma_{i,j}-\rho_{i,j})}{\partial \tau}\, .
\nonumber
\ee

Therefore, the total current conservation, 
\be
\nabla\cdot\left(\frac{\partial{\E}}{\partial \tau}+\sigma\E+D\nabla\sigma\right)=0\,,
\label{consistency}
\ee
also holds on the level of the discretizations.
                                                                                           
\subsection{Temporal discretization}
\label{sectime}
                                                                                           
After the spatial discretization, the system of equations (\ref{sds})
can be written in vector form as a system of ordinary differential equations,
\be
\begin{array}{lcr}
\displaystyle
\frac{d\sigma}{d\tau} & = & G(\sigma, {\E}) \, , \\[4mm]
\displaystyle
\frac{d\rho}{d\tau} & = & S(\sigma, {\E}) \, ,
\label{ode}
\end{array}
\ee
where the components of $G$ and $S$ are given by the spatial discretizations
in (\ref{sds}).  The electric field ${\E}$ and the
field strength $|{\E}|$ are computed from given $\sigma, \rho$
by discretized versions of (\ref{poiss2d}) and (\ref{compe}), discussed in Sect.~\ref{secrefpoiss}.
Therefore the full set of semi-discrete equations actually forms a system
of differential-algebraic equations.
                                                                                           
The particle densities are updated in time using the explicit trapezoidal
rule, which is a two-stage method, with step size $\Delta \tau$.
Starting at time $\tau^n=n\Delta \tau$ from known particle distributions
$\sigma^n(r,z) \approx \sigma(r,z,\tau^n)$, $\rho^n(r,z)\approx\rho(r,z,\tau^n)$
and known electric field
${\E}^n(r,z)\approx{\E}(r,z,\tau^n)$, the predictors
$\bar{\sigma}^{n+1}$ and $\bar{\rho}^{n+1}$ for the electron and ion
densities at time $\tau^{n+1}$ are first computed by
\be
\begin{array}{l}
\displaystyle
\bar{\sigma}^{n+1} \;=\; \sigma^n + \Delta \tau\, G(\sigma^n,{\E}^n) \, ,
\\[4mm]
\displaystyle
\bar{\rho}^{n+1} \;=\; \rho^n + \Delta \tau\, S(\sigma^n,{\E}^n)\, ,
\end{array}
\label{rk1}
\ee

After this the Poisson equation (\ref{poiss2d}) is solved with source
term $\bar{\sigma}^{n+1}-\bar{\rho}^{n+1}$, leading to the electric
field $\bar{\E}^{n+1}$ at this intermediate stage by
Eq.\,(\ref{compe}).
The final densities at the new time level $\tau^{n+1}$ are then computed as
\be
\label{timestep2}
\begin{array}{l}
\displaystyle
\sigma^{n+1} \;=\; \sigma^n
+ \frac{\Delta \tau}{2} G(\sigma^n,{\E}^n)
+ \frac{\Delta \tau}{2} G(\bar{\sigma}^{n+1},\bar{\E}^{n+1}) \, ,
\\[4mm]
\displaystyle
\rho^{n+1} \;=\; \rho^n
+ \frac{\Delta \tau}{2} S(\sigma^n,{\E}^n)
+ \frac{\Delta \tau}{2} S(\bar{\sigma}^{n+1},\bar{\E}^{n+1}) \, ,
\end{array}
\ee
after which the Poisson equation is solved once more, but now with the
source term $\sigma^{n+1}-\rho^{n+1}$, giving the electric field
${\E}^{n+1}$ induced by the final particle densities at time
$\tau^{n+1}$.
This time discretization is second-order accurate, which is in line with
the accuracy of the spatial discretization.

The use of an explicit time integration scheme implies that the grid spacing and
time step should obey restrictions for stability. For the advection part we get a 
Courant-Friedrichs-Lewy (CFL) restriction
\be
\displaystyle\max \mE_r \frac{\Delta\tau}{\Delta r}+\max \mE_z \frac{\Delta\tau}{\Delta z} & < & \nu_a\, ,
\label{CFL}
\ee
and the diffusion part leads to 
\be
\displaystyle D\frac{\Delta\tau}{\Delta r^2}+ D\frac{\Delta\tau}{\Delta z^2} & < & \nu_d\, .
\label{diffrestr}
\ee
Actually, to be more precise, a combination of~(\ref{CFL}),~(\ref{diffrestr}) should be considered. 
However, in our simulations, condition~(\ref{CFL}) will dominate and the time step will be chosen
well inside this constraint.

For the first order upwind advection scheme combined with a two-stage Runge-Kutta method,
the maximal Courant number is~\cite{hun2003} $\nu_a^1=1$, while for the third order upwind
scheme it is $\nu_a^3=0.87$. The second order central discretization demands a maximal
Courant number $\nu_d=0.5$.

A third restriction for the time step comes from the dielectric relaxation time. The ions
are considered to be immobile, which leaves us with the following time step
restriction in dimensional units~\cite{bar1986},
\be
\Delta t\le\frac{\epsilon_0}{e\mu_e\max n_e}\, ,
\ee
where we refer to the previous section for the meaning of these quantities. If we apply
the dimensional analysis of the previous section, we obtain the time step restriction in dimensionless units,
\be
\Delta\tau\le\frac{1}{\sigma}\, .
\label{DRT}
\ee
In practice, it appears that this  restriction is much weaker than that for the stability of the numerical scheme.
                                                                                           
The choice for an explicit time stepping scheme was made after tests with
VLUGR~\cite{blo1996}, a local refinement code that uses --computationally much
more intensive-- implicit schemes (BDF2).
These tests showed that these implicit schemes also needed relatively
small time steps to obtain accurate solutions, so that in the end an explicit
scheme would be more efficient in spite of stability restrictions for the
time steps. Moreover the use of an explicit scheme allows us to decouple
the computation of the particle densities from that of the electric potential
and electric fields. With a fully implicit scheme all quantities are
coupled, leading to much more complex computations and longer computation times.
                                                                                           
Another reason for preferring explicit time stepping is monotonicity.
The solutions in our model have very steep gradients for which we use spatial
discretizations with limiters to prevent spurious oscillations.
Of course, such oscillations should also be prevented in the time integration.
This has led to the development of schemes with TVD (total variation diminishing)
or SSP (strong stability preserving) properties; see~\cite{got2001,hun2003}.
In contrast to stability in the von Neumann sense (i.e., $L_2$-stability
for linear(ized) problems with (frozen) constant coefficients), such
monotonicity requirements do not allow large step sizes with implicit
methods of order larger than one.
Among the explicit methods, the explicit trapezoidal rule is optimal with
respect to such monotonicity requirements.

\subsection{Remarks on alternative discretizations}

The above combination of spatial and temporal discretizations
provide a relatively simple scheme for the streamer simulations.
The accuracy will be roughly $O(\Delta x^2) + O(\Delta \tau^2) $ in
regions where the solution is smooth (also for the limited advection
discretization~\cite[p.\,218]{hun2003}).
In our tests, step size restrictions mainly originate from the
advective parts in the continuity equations. The above scheme is
stable and monotone (free of oscillations) for Courant numbers up to
one, approximately. Usually we take smaller step sizes than imposed
by this bound to reduce temporal errors.

As mentioned before in Sect.\,\ref{subsecdiscrcont} using
a first-order upwind discretization for the advective term will usually give rise to too much
diffusion, whereas second-order central advection discretizations lead
to numerical oscillations and negative concentrations.
                                                                                           
Higher-order discretizations can certainly be viable alternatives.
However, we then will have larger spatial stencils, which creates
more difficulties with local grid refinements where numerical interfaces
are created. The above discretization is robust and easy to implement.
                                                                                           
It is well known that limiting as in (\ref{fluxa})
gives some clipping of peak values in linear advection tests, simply
because the limiter does not distinguish genuine extrema from
oscillations induced numerically.
This can be avoided by adjusting the limiter near extrema, but in
the streamer tests it was found that such adjustments were not
necessary. In the streamer model the local extrema in each spatial
direction are located in the streamer head, and the nonlinear
character of the equations gives a natural steepening there
which counteracts local numerical dissipation.
                                                                                           
In \cite{vit1994} a flux-corrected transport (FCT) scheme was used. The advantage of our semi-discrete
approach is that is becomes easier to add source terms without having to change
the simulation drastically. Also the transition from 1D
discretizations to 2D or 3D becomes straightforward conceptually;
the implementations for higher dimensions are still difficult, of course,
in particular for the Poisson equation.
Moreover, in~\cite{bob1998} comparisons of the FCT scheme with a scheme using the van Leer
limiter (which is closely related to the limiter used in our scheme) show
that the FCT scheme, in contrast to the limited scheme,  gives somewhat irregular results for simple advection
tests in regions with small densities. In the leading edge the densities 
decay exponentially, and we do not want such irregularities to occur.

\section{The adaptive refinement procedure}
\label{refsect}
\subsection{The limitations of the uniform grid approach}
\label{seclimuni}
                                                                                           
Up to now, all simulations that have been carried out on the minimal streamer
model were performed on a uniform grid~\cite{dha1987,vit1994,arr2002,roc2002}. 
However the use of a uniform grid on such a large system has several
limitations.  

The first limitation is the size of the system. In~\cite{arr2002,roc2002} the
simulations were performed in a radially symmetric geometry with
$2000\times2000=4\cdot10^6$ grid points. Since the number of variables is of at
least 10 per grid point (the electron and ion densities both at old and
new time step, the electric potential, the electric field components and
strength, and the terms containing the temporal derivatives of both electrons and ions),
the total number of variables will be at least $40\cdot 10^6$.
So, when computing in double precision (64 bits or 8-bytes values) the memory usage
is in the order of several hundreds MB, depending on the compiler.
Moreover, these simulations show that the streamer will eventually branch,
and up to now there was no convergence of the branching time with respect to the mesh size.
In order to investigate the branching, it would be necessary to rerun the
simulation with a smaller grid size.
Moreover it would be interesting to investigate larger systems. So from that
point of view it is worth looking at an algorithm that would require much
less memory usage.

The second limitation comes from the Poisson equation, which has to be solved
at each time step, and therefore requires a fast solver. 
For this we use the {\sc Fishpak} routine, described in~\cite{wes2001,schu1976}.
One of the major limitations of this routine is its ineptitude to deal
accurately with very large grids, due to numerical instabilities with
respect to round-off errors.
Numerical experiments show that on an $m_r \times m_z$ grid with either
$m_r$ or $m_z$ substantially larger than 2000, the {\sc Fishpak} routine
gives large errors due to numerical instabilities with respect to
round-off errors. An illustration is presented in the appendix.

It is necessary to develop some strategy to counteract these limitations.
This will be done by choosing separate grids with suitable local 
refinements for the continuity equations (\ref{sigmaeq})-(\ref{rhoeq}) 
and the Poisson equation (\ref{phieq}).

\subsection{General structure of the locally uniform refined grids}

Both the continuity equations (\ref{cdrs2d}) and the
Poisson equation (\ref{poiss2d}) are computed on a set of uniform,
radially symmetric grids, that are refined locally where needed.
Solving these equations separately rather than simultaneously
allows the use of different sets of grids for each equation, thereby making
it possible to decouple the grids for the continuity equation from those
of the Poisson equation; grids can then be tailored for the
particular task at hand. We emphasize that it is the use of an 
explicit time stepping method that allows us to decouple the grids.
                                                                                           
In both cases the equations are first solved on a coarse grid.
This grid is then refined in those regions where a refinement criterion
-- which will be treated in more detail below -- is met.
These finer grids can be further refined, wherever the criterion is still
satisfied.
Both the grids and the refinement criteria may be different for the
continuity equations on the one hand and the Poisson equation on the other
hand, but the general structure of the grids is the same for both type of
equations.
It consists of a series of nested grids $\Omega^k$, $k$ being the grid
number, with level $l(k)$.
This level function gives the mesh width of a grid, $l(1)=1$ corresponding
to the coarsest grid $\Omega^1$ with mesh width $\Delta r^c$ and $\Delta z^c$.
A certain grid will have a mesh twice as fine as the grid one level coarser --
which we will call its parent grid -- so that the mesh widths of a grid
with level $l$ become $\Delta r^l={\Delta r^c}/{2^{l-1}}$ and
$\Delta z^l={\Delta z^c}/{2^{l-1}}$.

\begin{figure}[t]
\begin{center}
\includegraphics[width=5cm]{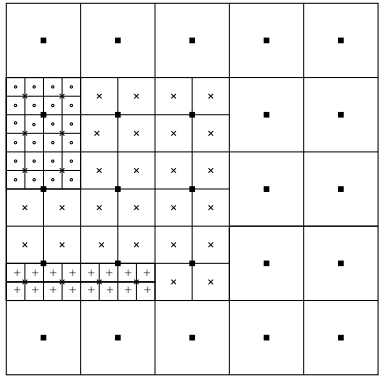}
\caption{\footnotesize\sf
An example of nested grids:
$\scriptscriptstyle\blacksquare$ cell centers on $\Omega^1$ with $l(1)=1$,
$\times$ cell centers on $\Omega^2$ with $l(2)=2$, $+$ cell centers on
$\Omega^3$ with $l(3)=3$,
$\circ$ cell centers on $\Omega^4$ with $l(4)=3$.}
\label{refgrid}
\end{center}
\end{figure}

Fig.\,\ref{refgrid} shows an example of four nested grids on three levels.
We will denote a quantity $u$ on a certain grid
$\Omega^k$ with level $l = l(k)$ as $u^{k}_{ij} = u(r^l_i,z^l_j)$.
All grids are characterized by the coordinates of their corners relative
to the origin of the system (on the axis of symmetry at the cathode), so 
that $r_i^l=(i-1/2)\Delta r^l$, $z_j^l=(j-1/2)\Delta z^l$.                                                                                           

Throughout this article, the grids for the Poisson equation are denoted by 
$\mathcal G$, those for the continuity equations by $\mathcal H$.  
For the continuity equations, the grids are structured as follows: we determine all grid cells
of a grid $\mc H^l$ at a certain level $l$ that, following some refinement monitor
(we will give more details later), have to be refined. This results in one or more 
sets of connected finer grid cells. As a first step, we chose the finer {\em child grids} $\{\mc H^{l+1}\}$ of 
the {\em parent grid} $\mc H^l$ to cover the smallest rectangular domain enclosing each of these sets
of connected grid cells.
Although this first rudimentary  approach gave some gain in computational time, it lead to a large 
number of unnecessary fine grid cells, due to the curved nature of the ionization front. 
So instead, we divide such unnecessary large rectangles into smaller rectangular patches, 
thereby limiting the number of fine grid cells. Obviously, the union of all these child 
grids $\{\mc H^{l+1}\}$ should contain all the grid cells indicated by the monitor. 
For programming reasons, the rectangular child grids $\{\mc H^{l+1}\}$ were chosen to 
all have an equal, arbitrary number $M_0$ of grid points in the radial 
direction. Using a large value for $M_0$ leads to unnecessary large child grids, using a
small value would make no sense because each grid requires the storage of boundary values.
Since the rectangular grid structure is preserved, this could be relatively easily be implemented 
in the existing code, and using a suitable value of $M_0$ lead to a significant gain in 
computational time and memory. 

To compute the electric potential, however, such a grid distribution is not appropriate. 
This comes from our use of a fast Poisson solver, which computes, on a rectangular grid,
the potential induced by a space charge distribution on that same rectangular grid.
The solution of the Poisson equation is not, however, determined locally, and these non
local effects are not accounted for properly if we compute the potential 
on smaller rectangles like the continuity equation. So in the case of the Poisson equation
we use the same grid structure as we first did for the continuity equation: we determine, 
using some refinement monitor, all grid cells of a grid $\mc G^l$ at a certain level $l$ that
have to be refined, and the finer {\em child grid} 
$\{\mc G^{l+1}\}$ of the {\em parent grid} $\mc G^l$ is set to cover the smallest rectangular 
domain enclosing all these grid cells. 

In what follows, to make the distinction between the indices on a coarse
and on a fine grid, we use capital indices $I$ and $J$ for the coarse
grid and small indices $i$ and $j$ for the fine grid.
Notice that a coarse grid cell with the cell center $(r_I,z_J)$ contains four
finer cells with centers $(r_i,z_j)$, $(r_i,z_{j+1})$, $(r_{i+1},z_j)$,
$(r_{i+1},z_{j+1})$, with $ i= 2I-1$ and $j=2J-1$.

Let us now discuss in more details the refinement algorithm for the continuity
equations and that for the Poisson equation, and the benefits of decoupling the
grids for both equations. 

\subsection{Refinement scheme for the continuity equation}
\label{secrefcont}

\begin{figure}[b]
\begin{center}
\mbox{\subfigure[]{\includegraphics[width=2.5cm]{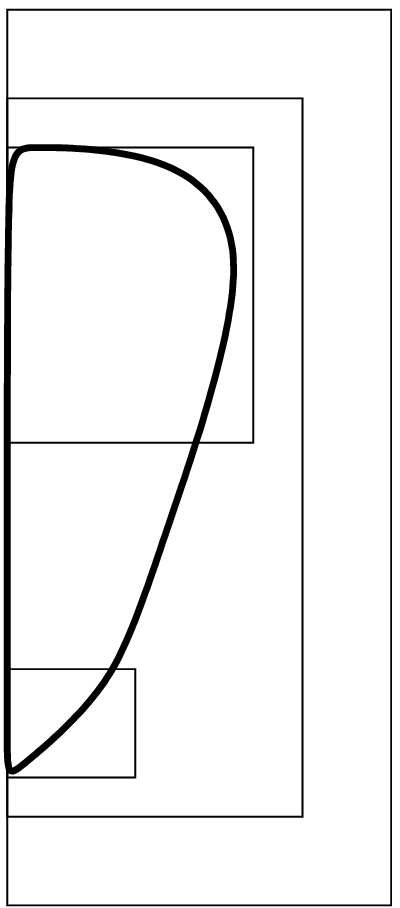}}\qquad
\subfigure[]{\includegraphics[width=2.5cm]{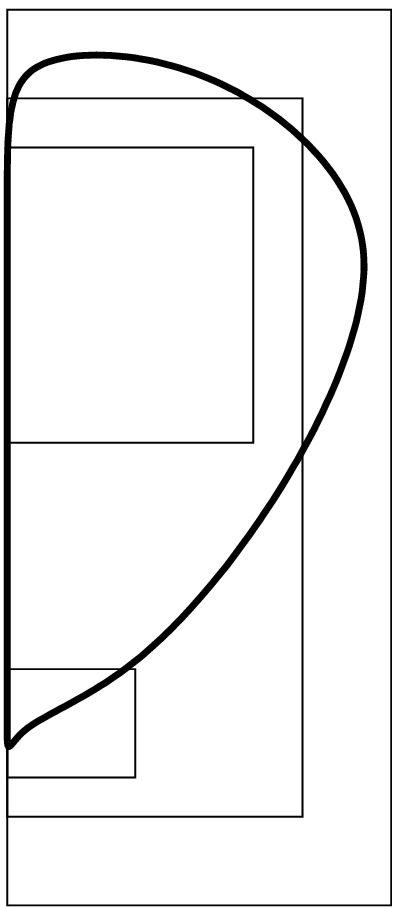}}\qquad
\subfigure[]{\includegraphics[width=3.cm]{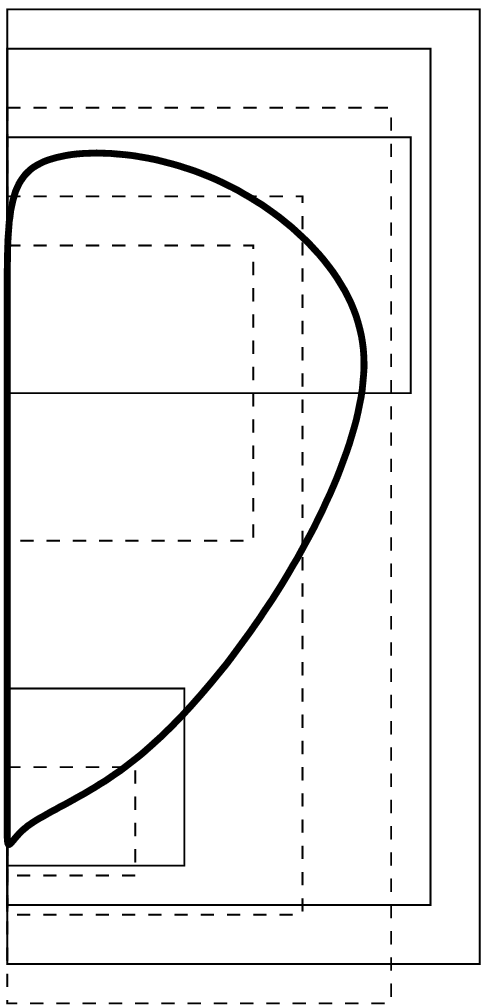}}}
\caption{\footnotesize\sf
(a): Contour line of the solution and the set of rectangular computational
grids, both at $\tau^n$.
(b): Contour line of the solution at $\tau^{n+1}$ and computational grids at $\tau^n$.
(c): Contour line and computational grids at $\tau^{n+1}$, the dashed lines are
the grids at $\tau^n$.}
\label{gridn}
\end{center}
\end{figure}

Let us assume that the particle distributions and electric field at time
$\tau^n$ are known on a set $S^n$ of $m$ rectangular grids $\mathcal H^{n,k}$ with
level $l=l(k)$, $1\leq k\leq m$, as shown in Fig.\,\ref{gridn}. 

Then, using the explicit time stepping method introduced in Sect.\,\ref{sectime},
the particle distributions at time $\tau^{n+1}$  can be computed on all the
grids of $S^n$ (Fig.\,\ref{gridn}b).
Now the new set $S^{n+1}$ of nested grids that is best suitable for the
solution at $\tau^{n+1}$ has to be found, as in Fig.\,\ref{gridn}c.

Moreover, the computational domain for the continuity equations can be reduced substantially 
by the following physical consideration: our model is a fluid model based in the continuum 
hypothesis, which is not valid anymore if the densities are below a certain threshold, that
is taken as 1 mm$^{-3}$. In nitrogen at atmospheric pressure, this corresponds roughly to 
a dimensionless density of 10$^{-12}$. The regions where all densities are below this threshold 
is ignored. Therefore, after each time step the densities below this threshold are set to zero. 
The computational domain for the continuity equations for the next time step is then taken as 
the region where $\sigma$ or $\rho$ are strictly positive. In view of our two-stage Runge-Kutta 
time stepping we use a four point extension of this domain in all directions.
                                                                                           
\subsubsection{Restriction of fine grid values to a coarse grid}
At first, when a grid $\mathcal H^{n,K}$ at level $L=L(K)$ contains a finer
grid $\mathcal H^{n,k}$ at level $l=L+1$, the particle densities on
$\mathcal H^{n,k}$ are restricted to $\mathcal H^{n,K}$ in such a way that the
total mass of each particle species is conserved locally.
From Fig.\,\ref{refgrid} it can be seen
that one cell of $\mc H^{n,K}$ contains four cells on $\mc H^{n,k}$,
and the mass conservation implies that for each particle species the
total mass in the coarse grid cell is equal to the sum of the masses in
the finer grid cells, which, taking into account the cylindrical geometry
of the cells, translates in the following restriction formula
$U^K = Res(U^k)$ for the grid functions $U^K = \{u^K_{IJ}\}$ and
$U^k = \{u^k_{ij}\}$,
\be
U^K_{IJ} = Res(U^k)_{IJ} =
\frac{1}{4r^L_I}\sum_{m=2I-1}^{2I}\sum_{n=2J-1}^{2J}r^{l}_mu^l_{mn} \, ,
\label{restr}
\ee
in which $u$ stands for either the electron or the ion density.
This restriction step is carried out because time stepping on a too coarse
grid may lead to erroneous values, which are now overwritten by the better
restricted values. The $r$-factors account for the mass distribution in the radial
cells in cylindrical symmetry. 
                                                                                           
\subsubsection{Refinement criterion: curvature monitor}
It is now possible to find the regions where the grids should be refined
at $t^{n+1}$.  The decision whether a finer grid should be used on a certain
region is made with a relative curvature monitor. This monitor is defined on
a grid with level $l$ as the discretization of
\be
M_u(r^l,z^l)=(\Delta r^l)^2\Big|\frac{1}{r}\frac{\partial}{\partial r}(r\frac{\partial u}{\partial r})\Big|+
(\Delta z^l)^2\Big|\frac{\partial^2u}{\partial z^2}\Big|\, .
\label{monitor}
\ee
Although this expression does not provide an accurate estimate of the discretization errors,
it does give a good indication of the degree of spatial difficulty of the
problem~\cite{tro1991}. It is applied first to the electron density $\sigma$,
since that is the quantity which advects and diffuses, and therefore the
quantity in which some discretization error will appear.
The monitor is also applied to the total charge density $\sigma-\rho$ since
this is the source term of the Poisson equation, and the accuracy of the
solution of the Poisson equation is of course dependent on the 
accuracy of the source term.
The monitor is taken relative to the maximum value of each quantity, and the refinement criterion 
through which the need to refine a certain grid $\mc H^k$ with level $l$ then reads: 
\be
\text{refine all grid cells $i,j$ where }\displaystyle\frac{M_u(r^l_i,z^l_j)}{\max u^l_{ij}} \geq \epsilon^l_u,\quad
u=\sigma,\,\sigma-\rho 
\label{refcrit}
\ee
in which $\epsilon^l_\sigma$ and $\epsilon^l_{\sigma-\rho}$ are grid-dependent
refinement tolerances that will further be discussed in Sect.~\ref{sec1D} and Sect.~\ref{sectest2d}.
                                                                                           
Now, starting from the coarsest grid, the monitors are computed by approximating
them with a second order central discretization, which determines the regions that should
be refined.
For this set of finer grids again the regions to be refined are computed, and
so on either until the finest discretization level is reached, or until the monitor
is small enough on every grid.
Now the new set of nested grids $S^{n+1}$ has been determined, but the particle
densities are still only known on the set $S^n$ (in Fig.\,\ref{gridn}c
this means that we have to convert the density distributions on the dashed
grids to distributions on the solid rectangles).

Criteria such as~(\ref{monitor})-(\ref{refcrit}) are common for grid refinements. As we shall see
in experiments, it will be necessary to extend the refined regions to include (a part of)
the leading edge of a streamer. This leading edge is the high field region into which the streamer propagates,
and where the densities decay exponentially. This modification, due to the pulled front
character of the equations~\cite{ebe2000}, is essential for the front dynamics to be well captured,
and is a major new insight for the simulation of streamers, and more generally, of any leading edge dominated problem.
It is discussed in more detail in Sect.\,\ref{sec1D}.
                                                                                           
\begin{figure}[htb]
\begin{center}
\includegraphics[width=6cm]{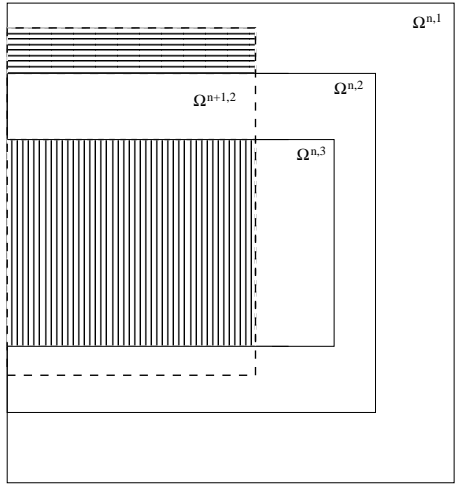}
\caption{\footnotesize\sf
Three grids at time $\tau^n$ (solid lines) and a grid at next time step $\tau^{n+1}$
(enclosed by the dashed lines) with the same level as $\mc H^{n,2}$.
In the vertically striped region the new grid coincides with a finer
grid at the previous time step, in the horizontally striped region
only a coarser grid existed at $\tau^n$, and in the region that is not
filled a grid at the same level existed at $\tau^n$.}
\label{gridnn+1}
\end{center}
\end{figure}

\subsubsection{Grid mapping}
The refinement monitor gives a new grid distribution $\{\mc H^{n+1,k}\}$ at time
$\tau^{n+1}$. In the following  we consider mapping functions used to
map the solution at $\tau^{n+1}$ on the previous grid distribution $\{\mc H^{n,k}\}$ 
on the new grid distribution $\{\mc H^{n+1,k}\}$.
There are three possible relations between a grid $\mc H^{n+1,k}$ at time
$\tau^{n+1}$ and the older grids $\{\mc H^{n,k}\}$, as illustrated in
Fig.\,\ref{gridnn+1}.

First it is possible that a fine grid $\mc H^{n,k}$ at the previous time step
now is covered by a coarser grid $\mc H^{n+1,K}$ (as in the vertically striped
region of Fig.\,\ref{gridnn+1}).  Then the values of the densities on the
new grid are computed by the restriction~(\ref{restr}):
\be
U^{n+1,K} = Res(U^{n,k}) \quad \text{ on } \mc H^{n+1,K}\cap\mc H^{n,k}\, ,
\ee
where $U$ again stands for the set of grid values of either electron- or
ion densities.

Secondly, there is the possibility that (part of) the new grid already
existed at previous time step, in which case there is no need for projecting
the density distributions from one grid to the other.

Finally, it may occur that the new grid lies on a region where only
a coarser grid $\mc H^{n,K}$ existed at the previous time level (as in
the horizontally striped region in Fig.\,\ref{gridnn+1}).
Then we have to prolongate the coarse grid values $U^{n+1,K}$
to the new fine grid. One main consideration in the choice of the prolongation
is the conservation of charge in the discretizations. For simplicity, we will first consider 
a one-dimensional prolongation, which is then easily extended to more dimensions.
We consider 
coarse grid values $U^K=\{u^K_{I}\}$. Then a mass conserving interpolation
for values $U^k = \{u^k_{i}\}$  on a grid twice as fine is,
\be
u^k_i=U^K_I+D_I  \, ,\quad u^k_{i+1}=U^K_I-D_I\, ,
\ee
which obviously implies mass conservation, $\Delta z^lu^k_i+\Delta z^lu^k_{i+1}=\Delta z^LU^K_I$.
Using a three-points stencil, the coefficient $D_I$, such that the interpolation is
second order accurate, can be written as
\be
D_I=\frac{1}{8}(U_{I-1}+U_{I+1})\, .
\ee 

In the case of a three-dimensional geometry with axial symmetry, the above interpolation is applied
on  $U^K$ in the $z$-direction, and on $RU^K$ in the radial direction, which ensures mass conservation on
such a system.
\\
\noindent{\em Remark}: This mass conserving interpolation might lead to new extrema or negative
values. That could be prevented automatically by limiting the size of the slopes. However,
in our simulations this turned out not to be necessary.

Finally, we need boundary values for all grids. On the coarsest grid these simply
follow from the discretization of the boundary conditions (\ref{bcphi})-(\ref{bcsigma}), as
explained in Sect.\,\ref{subsecdiscrcont}. On finer grids they
are interpolated bi-quadratically from coarse grid values. The interpolation error is then third
order, and therefore smaller than the discretization error, which would not be the case with a bi-linear
interpolation. 
The quadratic interpolation is derived using the Lagrange interpolation formula 
to find $U^k = \{u^k_{ij}\}$ from
the coarse grid values $U^K=\{u^K_{IJ}\}$. The interpolated value becomes
\be
\label{int}
u^k_{ij} =  Int(U^K)_{ij}
 =  \displaystyle\frac{1}{r^{l(k)}}\sum_{p=-1}^1\sum_{q=-1}^1r^{l(K)}_{I+p}u^K_{I+p,J+q}
\prod_{\substack{P\neq p\\P=-1}}^1
\frac{r^{l(k)}_i-r^{l(K)}_{I+P}}{r^{l(K)}_{I+p}-r^{l(K)}_{I+P}}
\prod_{\substack{Q\neq q\\Q=-1}}^1 
\frac{z^{l(k)}_j-z^{l(K)}_{J+Q}}{z^{l(K)}_{J+q}-z^{l(K)}_{J+Q}}\, ,
\label{quadinteq}
\ee
in which $I=(i+i\mod 2)/2$ and $J=(j+j\mod 2)/2$. 
We notice that the interpolated quantity again is not $\sigma$ but $r\sigma$ because
of the cylindrical coordinate system.

In our algorithm, mass conservation 
at the grid interfaces will simply be ensured by matching the fluxes
of the fine and coarse grids.

\subsubsection{Flux conservation at grid interfaces}

The mapping from one grid to the other must take into account 
a flux correction on grid interfaces, in order to ensure mass conservation. This
correction yields that the total mass going through one coarse grid cell must
be the same as the sum of total mass going through two fine grid cells.
Taking into account the cylindrical geometry of the system, the fluxes through coarse grid
cells become, in terms of the fine grid fluxes, 
\be
\begin{array}{rcl}
F^K_{r;I+1/2,J} & = & \displaystyle\frac{\Delta r^l}{\Delta r^L}(F^k_{r;i+1/2,j}+F^k_{r;i+1/2,j+1})\, , \\
\displaystyle F^K_{z;I,J+1/2} & = & 
      \displaystyle \frac{\Delta r^l}{\Delta r^Lr_I}(r^l_iF^k_{z;i,j+1/2}+r^l_{i+1}F^k_{z;i+1,j+1/2})\, , 
\end{array}
\label{fluxcorrz}
\ee
where $i=2I-1$ and $j=2J-1$.

\subsection{Refinement scheme for the Poisson equation}
\label{secrefpoiss}

The refinement strategy for the computation of the electric potential 
is also based on a static regridding approach, where
the grids are adapted after each complete time step.
In this case however, the refinement criterion is not based on a curvature
monitor but on an iterative error estimate approach.
The {\sc Fishpak} routine will be used on a sequence of nested grids
${\mathcal G}^m$.
The solution on a coarse grid will be used to provide boundary conditions
for the grid on the finer level, which will in general extend over a smaller
region.
This approach is explained in full detail in~\cite{wac2005}; here we will discuss
the main features of the scheme.

Starting from the (known) charge density distribution $\sigma-\rho$ on a set
of grids $\{\mc H^k\}$, the Poisson equation is first solved on the two
coarsest grids ${\mathcal G}^1$ and ${\mathcal G}^2$, both covering the
entire computational domain $(0,L_r)\times(0,L_z)$. 
The finest of these two grids is coarser or as coarse as the coarsest grid
of $\{\mc H^k\}$. The densities should then first be mapped onto
the coarse grids ${\mathcal G}^1$ and ${\mathcal G}^2$, using the restriction formula (\ref{restr}).
The source term of the Poisson equation is then known on these two coarse
grids, and the Poisson equation is solved using a {\sc Fishpak} routine
that discretizes Eq.\,(\ref{poiss2d}) using second-order differences:
\be
\sigma^m_{i,j}-\rho^m_{i,j} =
\frac{\phi^m_{i+1,j}-2\phi^m_{i,j}+\phi^m_{i-1,j}}{(\Delta r^{l})^2}
+ \frac{\phi^m_{i+1,j}-\phi^m_{i-1,j}}{2r^{l}_i\Delta r^{l}}
+ \frac{\phi^m_{i,j+1}-\phi^m_{i,j}+\phi^m_{i,j-1}}{(\Delta z^{l})^2}
\label{pdics}
\ee
in which $l = l(m)$ is the level of grid ${\mathcal G}^m$.
This system of linear equations is then solved using a cyclic reduction
algorithm, see~\cite{gol1996}. The details of {\sc Fishpak} are described
in~\cite{schu1976}. The subroutine was used as a black box in our simulations.
A comparison with iterative solvers, multigrid or conjugate gradient type,
can be found in~\cite{bot1997}. For the special problem we have here -- Poisson
equation on a rectangle -- such iterative solvers are not only much slower than
{\sc Fishpak}, but they also require much more computational memory.
                                                                                           
\begin{figure}[tb]
\begin{center}
\includegraphics[width=5cm]{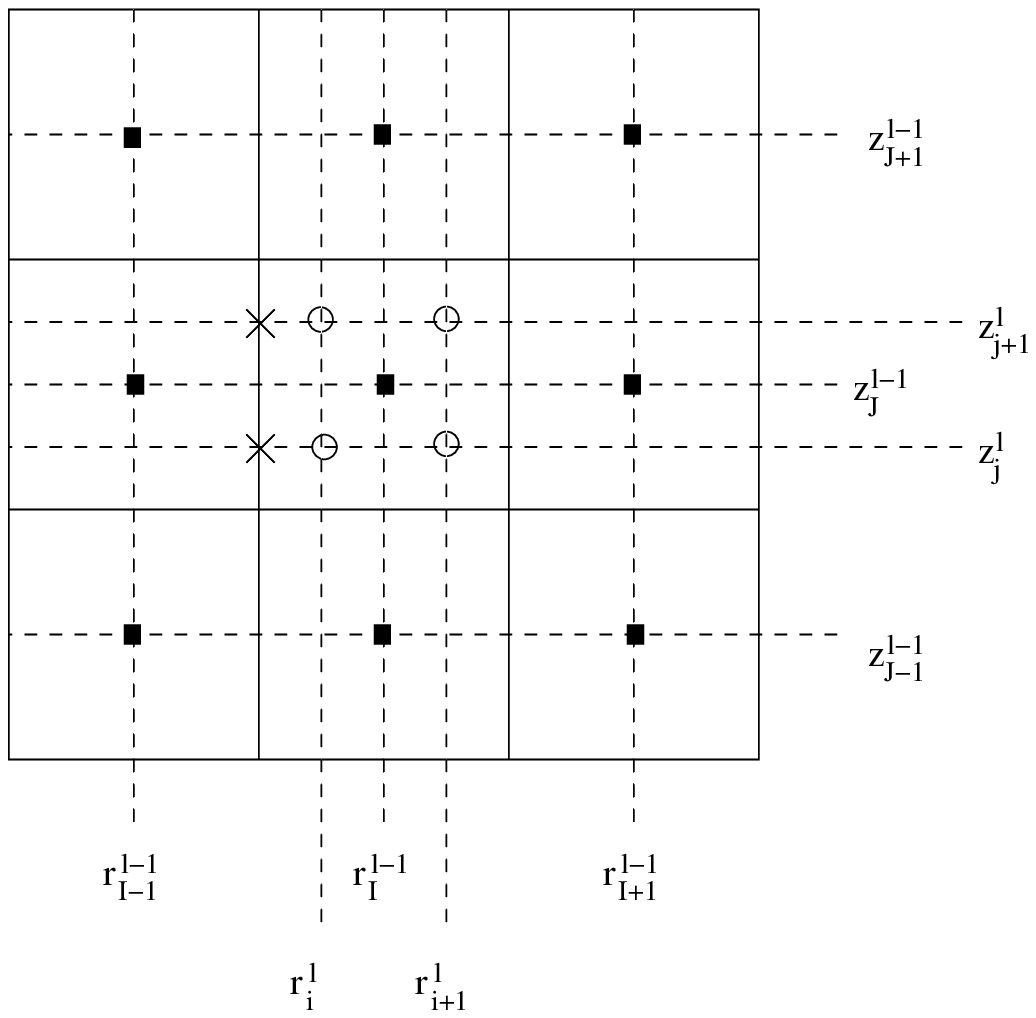}
\caption{\footnotesize\sf
The nine-points stencil used to map coarse grid values of the electric
potential $\phi^{l-1}_{IJ}$ onto the fine grid, thus obtaining $\phi^l_{i,j}$,
$\phi^l_{i+1,j}$, $\phi^l_{i,j+1}$ and $\phi^l_{i+1,j+1}$. The cell centers
of the coarse grid are marked with $\scriptscriptstyle\blacksquare$, those
of the fine grids with $\circ$, and the boundary points of the fine grids
are marked with $\times$.}
\label{restrp}
\end{center}
\end{figure}

As a next step, the coarse grid electric potentials $\phi^1$ on ${\mathcal G}^1$
and $\phi^2$ on ${\mathcal G}^2$ are compared with each other, by mapping
$\phi^1$ onto ${\mathcal G}^2$ with a quadratic interpolation based on a
nine-point stencil as shown in Fig.\,\ref{restrp}.
                                                                                           
For the prolongation that gives a continuous map of an electric potential
$\phi^M = \{\phi^M_{IJ}\}$ onto a finer grid ${\mathcal G}^m$ the
following formula, based on a least square fit, is used:
\be
\begin{array}{lcl}
\phi^m(r,z) & = &  Pro(\;\phi^M\,(r,z))
\\[2mm]
& = & \phi^{M}_{IJ} + c_{10}(r-r^{M}_I) + c_{01}(z-z^{M}_J) + 
c_{11}(r-r^{M}_I)(z-z^{M}_J) +\\[2mm]
& + & \displaystyle c_{20}(r-r^{M}_I)^2 + c_{02}(z-z^{l-1}_J)^2\, .
\label{intp}
\end{array}
\ee
For the values of the coefficients $c_{ij}$ we refer to~\cite{wac2005}.
This interpolation will primarily be used to generate boundary conditions
for the finer level, as illustrated in Fig.\,\ref{restrp} for the boundary
points on the finer grid marked with~$\times$, but it will also be used for
error estimation.
The interpolation for the boundary conditions will be done such that
there is a bias in the stencil toward the smooth region to enhance
the accuracy.
                                                                                           
The error of the solution on a grid $\mc G^m$ is then estimated by the
point-wise difference between the solution on that grid and the interpolation
of the solution on the underlying coarser grid $\mc G^M$:
\begin{equation}
\tilde g^m_{i,j} = \phi^m_{i,j} - (Pro\;\phi^M)_{i,j} \, .
\label{interrp}
\end{equation}
The refinement criterion is as follows:
\begin{equation}
\text{refine all grid cells $i,j$ where} \;\; |\tilde g^m_{i,j}|\ge \epsilon_\phi
\label{refcritp}
\end{equation}
where $\epsilon_\phi$ is a small parameter that still has to be chosen.
This leads to new and finer grids on which the whole procedure of mapping
the charge densities onto it, solving the Poisson equation using the
{\sc Fishpak} routine and estimating the error is repeated, and so on either until
the error is smaller than $\epsilon_\phi$ everywhere or until the finest
desired level is reached.
Notice that, since the grids do not cover the whole computational domain
anymore, the boundary conditions (\ref{bcphi}) do no longer hold.
We now have Dirichlet boundary conditions on each boundary that lies in
the interior of the computational domain, and they are computed by
interpolating the electric potential on the finest grid that is known using
Eq.\,(\ref{intp}). Using this third order interpolation formula implies
that the error introduced by the interpolation is negligible compared to
to discretization error, which is second order.
With this method there is an upper bound for the maximum error $e^m$ 
on grid $m$ with level $l$ \cite{wac2005}:
\begin{equation}
e^m\leq 3\max_{i,j}\tilde g^m_{ij}+(l-1)\epsilon_\phi,
\label{errphi}
\end{equation}
where $\tilde g^m=\max \tilde g_{ij}^m$ as defined in Eq.\,(\ref{interrp}). This means that 
the extra error due to the refinement can be made as small as wanted 
provided $\epsilon_\phi$ is taken small enough. Therefore, iteration,
e.g. with defect corrections, is not needed. Inequality~(\ref{errphi}) is
based on the assumption that the interpolation errors are negligible 
compared to the second-order discretization errors of the local problems,
and therefore interpolant~(\ref{intp}) was chosen to be of higher order. 
Although tacit smoothness assumptions are involved here, tests
in~\cite{wac2005} with strong local source terms did show
that the errors are indeed well controlled by this nested procedure.

We notice that this global error control is the reason for the choice  
of this particular refinement monitor for the Poisson equation. Such an
error control does not hold for the continuity equations, for which
we use the more suitable curvature monitor~\cite{blo1996,tro1991}. 

The electric field has to be known on the continuity grids ${\mc H^k}$,
since it appears in the continuity equations (\ref{sigmaeq})-(\ref{rhoeq}).
However it has to be computed from the electric potential that is only
known on the Poisson grids, using Eq.\,(\ref{compe}).
We consider the grid $\mc H^k$ with level $l(k)$ on which the
electric field has to be known, and the finest potential grid $\mc G^m$ with level
$l(m)$, which has a non-empty intersection with $\mc H^k$.

There are three possible situations: the potential grid can be either
coarser, as fine as, or finer than the continuity grid.
If both grids have the same size ($l(k)=l(m)$), or if the continuity grid
is coarser than the Poisson grid ($l(k)<l(m)$), then the field components
are set directly with a second order central approximation of Eq.\,(\ref{compe}),
using the neighboring points of the point where the field components need
to be known.
If the Poisson grid is coarser than the continuity grid ($l(k)>l(m)$), the
electric field is first computed on the Poisson grid with a second order
central approximation of Eq.\,(\ref{compe}), and then this field is
interpolated to the continuity grid.
The interpolation is performed with a piecewise bilinear approximation.
                                                                                           
\subsection{Overall algorithm} 
In previous sections, the refinement algorithms for the continuity equations and the Poisson
equation have been treated separately. We here describe the overall algorithm.

We start from known density distributions of the charged particles $\sigma^n$ and 
$\rho^n$ at a certain time $\tau^n$, on a set of grids ${\mathcal H^{n,k}}$. The electric
field induced by the charges on the grids ${\mathcal H^{n,k}}$ is computed
using the refinement method described in Sect.\,\ref{secrefpoiss}.

The step size for the time integration is then set in such a way that the stability 
conditions~(\ref{CFL})-(\ref{diffrestr}) of both advection and diffusion discretizations 
are met on the finest grid.
The values of both $\nu_a$ and $\nu_d$ have been taken as 0.1. This is smaller than the 
maximal values of 0.87 and 0.37 specific to the third-order
upwind scheme for the advection and the second-order central discretization for diffusion,
respectively, together with a two-stage Runge Kutta time integration method~\cite{hun2003}.

Then, starting on the finest grid level, the particle fluxes are computed on each grid  
and corrected using the flux correction formulas (\ref{fluxcorrz}). 
Then the first stage of the Runge Kutta step (\ref{rk1}) is carried out. The field induced 
by the density predictors $\bar{\sigma}^n$ and $\bar{\rho}^n$ is again computed using the procedure described in 
Sect.\,\ref{secrefpoiss}. The new boundary conditions for the particle densities are computed on
all grids  ${\mathcal H^{n,k}}$, and the second stage of the Runga-Kutta method is carried out in the 
same way as the first stage. We then obtain the density distributions $\sigma^{n+1}$ and
$\rho^{n+1}$ on the set of grids ${\mathcal H^{n,k}}$. 

Following the procedure described in Sect.\,\ref{secrefcont}, after restriction of fine grid values
to the parent grids, the grids ${\mathcal H^{n+1,k}}$ for the next time step are computed using the 
refinement monitor (\ref{refcrit}). The densities are then mapped from the grids  ${\mathcal H^{n,k}}$
to the grids  ${\mathcal H^{n+1,k}}$, and the new boundary conditions are computed. We thus
obtain the density distributions  $\sigma^{n+1}$ and $\rho^{n+1}$ on
the set of grids ${\mathcal H^{n+1,k}}$ at the new time $\tau^{n+1}$.

\subsection{Relations with other refinement algorithms}

As mentioned before, the initial attempts to solve the system~(\ref{cdrs2d})-(\ref{compe}) 
were done by using the existing adaptive finite difference code VLUGR~\cite{blo1996}. 
This code failed for our problem. Another unsuccessful attempt was
made using the adaptive finite element code KARDOS~\cite{La00}.
Both these codes use implicit time stepping, and they do not take 
into account the unstable behavior of the solution ahead of the front.

Since our algorithm uses explicit (Runge-Kutta) time stepping, the
refinement procedure is relatively simple. In particular, the computation
for the electric potential becomes decoupled from the density updates
in the continuity equations. This allows for a tailored approach to
these separate problems. The potential updates are performed using 
nested fast Poisson solvers~\cite{wac2005}, which requires rectangular
(nested) regions for these sub-problems together with a high-order
interpolant for a global error indicator.
For the continuity equations, on the other hand, the computational regions
can be chosen quite small, essentially confined to the streamer itself.
For these equation the simple error monitor based on local curvature 
performs well. 

Local time stepping, for the different grids on parts of the domain, 
have not been considered in this paper. Although savings might be expected 
for the continuity equations, where we now use the same time step dictated 
by the finest mesh, such approach would lead to the complication that 
also the potential needs to be updated locally, and that charge conservation
is no longer straightforward.  

A grid refinement approach with fixed, a priori chosen, grids but 
with local time stepping has been discussed in~\cite{CDW99b} for a, 
somewhat related, system of plasma equations. 
The main difference with our problem is that this system is considered 
with low electric fields but on a much larger time scale, with movement 
of ions described by Euler equations.
The time step restriction~(\ref{DRT}) for dielectric relaxation then 
becomes very severe. To overcome this restriction, an implicit treatment 
of the potential in the electron drift fluxes is required. 
In the approach of~\cite{CDW99a,CDW99b} this is done through~(\ref{consistency}) 
together with a low-order prediction update for the electron densities.
The other processes, including higher-order corrections, are solved in 
a time-split fashion. With local time steps, synchronization at the new 
(global) time level is needed to ensure the relevant conservation 
properties to hold, such as charge conservation.
Apparently, the instabilities in the leading edge are not that much of
a problem on the time scales considered in these papers, probably because
the charged fronts are smoother.

Some results based on non-uniform, moving grids have been reported~\cite{pan2003},
for the simulations of streamers originating from point electrodes. 
The regridding approach used in~\cite{pan2003} allows only for a fixed number of 
grid points, and does not seem to take into account properly the leading edge. 
In contrast, the method presented in this article enables a a fine grid to be 
put wherever needed, in particular, over the leading edge. 

We are not aware of other grid refinement approaches for systems of plasma 
equations that do take this leading edge instability into account.  Related 
problems have been reported however in~\cite{LPR98} with moving mesh 
methods for the 1D Fisher equation $u_t = u_{xx} + \gamma u (1 - u)$.
To overcome instability of the numerical moving mesh scheme, a special
monitor function was advised in~\cite{QS98}. This has essentially the 
effect of mesh refinement ahead of the front. It seems that extensions 
of this moving mesh approach to multidimensional reaction-diffusion 
equations or more complicated systems have not been implemented yet.

\section{Tests of the algorithm on a planar ionization front}
\label{sec1D}                                 

The implementation of the grid refinements is first tested on
the evolution of a planar streamer front. As analytical results
for this case are available~\cite{ebe1997}, any errors
in the implementation can be identified. Moreover, conclusions on 
the tolerances and interpolations can be drawn for the more general 
two-dimensional simulations. 

A one-dimensional simulation is carried out with the two-dimensional code
by letting initial and boundary conditions depend on $z$ only and not
on the radial coordinate $r$. Specifically, the initial ionization seed
\begin{equation}
\sigma(r,z,0) \;=\; \rho(r,z,0) \;=\; 10^{-2}\;e^{-(z-z_b)^2}
\label{ic1d}
\end{equation}
is located at $z=z_b$ and a constant background electric field 
${\E_b}=-|\E_b|{\bf\hat{e}_z}$ is applied.
The spatial region in this specific example is $(r,z)\in[0,L]\times[0,L]$ 
with $L=1024$. The boundary conditions for the electron density 
and the electric potential in this specific case are
\begin{equation}
\begin{array}{l}
\displaystyle
\sigma(r,0,\tau) = 0 \, , \quad
\frac{\partial\sigma}{\partial z}(r,L,\tau) = 0\, ,
\quad
\frac{\partial\sigma}{\partial r}(0,z,\tau) = 0\, , \quad
\frac{\partial\sigma}{\partial r}(L,z,\tau) = 0\, ,
\\[4mm]
\displaystyle
\phi(r,0,\tau) = 0 \, , \quad
\frac{\partial\phi}{\partial z}(r,L,\tau) = |\E_b| \, , \quad
\frac{\partial\phi}{\partial r}(0,z,\tau) = 0 \, , \quad
\frac{\partial\phi}{\partial r}(L,z,\tau) = 0 \, .
\end{array}
\end{equation}

In this situation, the electric field does not depend on $r$. 
It can be written as ${\bs{\mc E}}(r,z,\tau)={\mc E}(z,\tau){\bf \hat{e}_z}$,
and it can be obtained directly from the charge densities 
by integrating $\nabla\cdot{\bs{\mc E}}=\rho-\sigma$ from Eq.\,(\ref{phieq}) 
along the $z$-direction and using the boundary condition 
${\mc E} = -|\E_b|$ for the electric field at $z=L$. The result is
\begin{equation}
{{\mc E}}(z,\tau) = -|\E_b| 
+ \int_z^L\left(\sigma(z',\tau)-\rho(z',\tau)\right)\: dz'.
\label{Eint}
\end{equation}
This means that it is not necessary to calculate the electric potential $\phi$.
Rather the electron and positive ion density at time $t^n$ determine 
the electric field  ${\mc E}^n$ at each cell vertex by discretizing
Eq.~(\ref{Eint}). Starting from the value at $z=L$, which corresponds 
to $j=M$ on a grid with $M$ grid points in the $z$-direction, we thus obtain:
\begin{equation}
{{\mc E}}^n_{M+\frac{1}{2}} = -|\E_b| \,,\qquad
{{\mc E}}^n_{j-\frac{1}{2}} = {{\mc E}}^n_{j+\frac{1}{2}} 
+ \Delta z (\sigma^n_{j}-\rho^n_{j}) \;\; {\rm for}~ j \le M \,.
\label{e1d}
\end{equation}
The electric field strength in the cell centers is then taken as the
average field of the corresponding vertices,
\begin{equation}
|{\mc E}|^n_j = \frac{1}{2}\left|\mE^n_{j-\frac{1}{2}}
+ \mE^n_{j+\frac{1}{2}} \right| \label{e1d1}
\end{equation}
We emphasize that in this section, in which the particular case of a one-dimensional
streamer front is considered, we use Eqs.~(\ref{e1d})-(\ref{e1d1}) to compute the
electric field, rather than the Poisson equation. This speeds up considerably 
the computations, and enables us to test the
refinements for the continuity equations only. For tests on the refinements of the
Poisson equation only we refer to~\cite{wac2005}.

\begin{figure}
\begin{center}
\includegraphics[width=10cm]{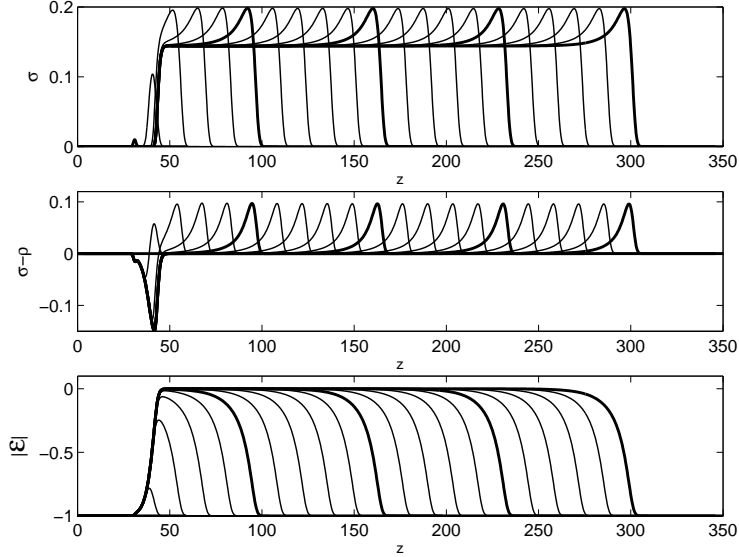}
\caption{\footnotesize\sf Numerical solutions of the planar streamer 
front with $|\E_b|=1$ and $D=0.1$ for $0\le\tau\le200$
with time steps of 10. The solution is computed on a uniform grid
with a mesh spacing of $1/4$ in the direction of propagation. The thick lines correspond to  multiples of 50
for $\tau$.}
\label{timeevol}
\end{center}
\end{figure}

Fig.~\ref{timeevol} shows the temporal evolution of the initial 
ionization seed (\ref{ic1d}) located at $z_b=31$, with a background 
field $|\E_b|=1$ and a diffusion coefficient $D=0.1$. The results shown 
are obtained on a uniform grid with mesh spacings $\Delta r=32$ and $\Delta z=1/4$,
which will be considered afterwards as a reference solution. 
$z_b$ is chosen large enough that boundary effects at $z=0$ do not matter.
and that the initial maximum is well presented even in the case treated later in
which a coarse grid with mesh size $\Delta z=2$ is used. 
After an initial growth until time $\tau\approx25$, 
an ionization front emerges that moves with about constant velocity
to the right in the direction opposite to the background field. 

\begin{figure} \begin{center}
\includegraphics[width=10cm]{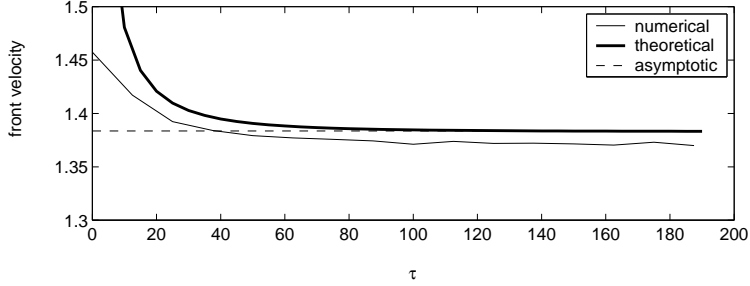}
\caption{\footnotesize\sf Numerical front velocity (thin solid line) as a function of time, compared
to the theoretical front velocity (thick line, derived in Appendix B) and the theoretical asymptotic front velocity $v^\ast=1.3863$} 
\label{vfrontf}
\end{center}
\end{figure}

The front velocity as a function of time is plotted in Fig.~\ref{vfrontf}. 
It is derived from the numerical displacement of the level $\sigma_f=10^{-8}$ 
within a time interval of 10. The front decelerates and eventually 
approaches a value somewhat below the asymptotic front 
velocity~\cite{ebe1997}
\begin{equation}
v^\ast=|\E_b|+2\sqrt{D|\E_b|e^{-1/|\E_b|}}
\label{asvel}
\end{equation}
which for these particular values of the background electric field and 
the diffusion coefficient is equal to 1.3836. The numerical velocity at
large times is around 1.365 which corresponds within an error margin 
of 1.5\% to the asymtpotic value. We expect to obtain even better results 
on finer grids, but we focus in what follows on the performance 
of the refinement algorithm compared to this uniform grid computation. For further
discussion of the results we refer to Appendix B. Moreover, deviations from this
asymptotic value can be derived analytically for different
numerical schemes, and that this is illustrated in Section 5.6.6
of~\cite{ebe2000} for an explicit and a semi-implicit time discretization
of a diffusion equation.

In the following illustrations we have computed the temporal evolution of the
densities and the electric field on a fine ($\Delta z=1/4$) and on a coarse 
($\Delta z=2$) uniform grid as well as on locally refined grids. In the latter case, 
the coarsest grid has a mesh spacing of $\Delta z^c=2$ which we refine 
up to a finest mesh width of $\Delta z^f=1/4$, thereby allowing for three levels of
refinement. The electric field is again computed using Eq.~\ref{e1d} rather than through
the Poisson equation for the electric potential, which speeds up the computations.  
The refinement algorithm for the continuity equations is as explained in 
Sect.\,\ref{secrefcont}.
To demonstrate that the leading edge has to be included in the
refinement, results with the ``standard'' refinement (i.e. without including the leading
edge)  will be compared to those that do include the leading edge.
In this leading edge of the ionization front the densities decay exponentially,
but the electric field strength is such that the reaction term is non-negligible.
From theoretical studies~\cite{ebe1997} as briefly recalled in Appendix B 
it is known that this region is
very important since it determines the asymptotic dynamics of the front.

For the present problem we use the a priori knowledge that the front moves
to the right and the leading edge is the region ahead of the front.
The standard criterion reads,
\begin{equation}
\text{refine all grid cells $j$ where} \;\;
\Big|\frac{\partial^2u}{\partial z^2}(z_j)\Big| \geq \epsilon_u ,\quad u=\sigma,\,\sigma-\rho
\label{refcrit1}
\end{equation}

Second, we use the same criterion but now with the inclusion of the
so-called leading edge in the refined region, taking into account
the cut-off of densities below the continuum threshold of 10$^{-12}$, i.e.:
\begin{equation}
\begin{array}{l}
\text{refine all grid cells $j$ obeying criterion (\ref{refcrit1})}, \\
\text{extend the refined region in the propagation direction} \\
\text{to all $\{z_j|\sigma_j>10^{-12}\}$.}
\label{refcritle}
\end{array}
\end{equation}
                                                                                           
\begin{figure}
\begin{center}
\includegraphics[width=10cm]{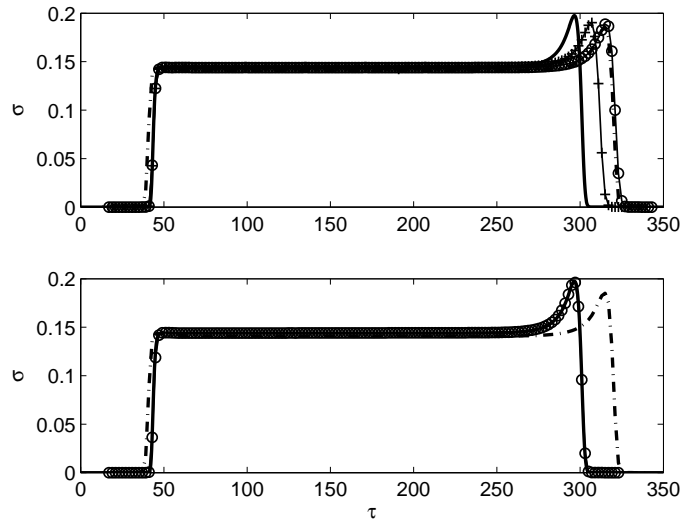}
\caption{\footnotesize\sf
Performance of the refinements without and with inclusion of the leading edge. 
Plotted is $\sigma(z,\tau)$ as a function of $z$ for $\tau=200$.
{\em Upper row}: results on grids refined with the standard criterion 
with a tolerance of $10^{-3}$ (circles) and $10^{-8}$ (pluses) compared 
to a fine grid computation ($\Delta z=\frac{1}{4}$, thick solid line) 
and a coarse grid computation ($\Delta z=2$, thick dash dotted line). 
{\em Lower row}: the same, but now with leading edge included 
in the refinement, only shown with a tolerance of $10^{-3}$.}
\label{res1d}
\end{center}
\end{figure}
                                                                              
The upper and lower plot of Fig.\,\ref{res1d} show the results 
of the one-dimensional streamer
simulation computed with the criterion~\ref{refcrit1} and~\ref{refcritle}, respectively,
together with a coarse and a fine uniform grid computation.
The  time step is such that the Courant numbers for both
advection and diffusion are at most 0.1, which is sufficiently
small to render the temporal errors negligible.

The value of $z_b$ in the initial pulse (\ref{ic1d}) has been chosen such
that the maximum of this Gaussian pulse was situated in a coarse grid 
cell center. Moreover, if a grid refinement is needed at the very first 
time step, the initial condition on the finer grids is not interpolated 
from the coarser grid values, but is calculated directly from 
Eq.~(\ref{ic1d}), so that the initial pulse on the finer grids 
is computed without interpolation errors.

The results with a fine uniform grid can be considered as a reference solution,
and it is clear that on the coarse uniform grid the front moves too fast and
is too smooth.  This is due to the large amount of
numerical diffusion introduced by the coarse grid.
As can be concluded directly from the expression for 
the asymptotic front velocity, a larger diffusion constant leads to
a larger velocity, and a larger velocity makes the front smoother
\cite{ebe1997}.

The same is observed on the grids refined according to the standard criterion,
even with a very strict tolerance the front is badly captured. This also appeared 
with other standard refinements codes, e.g. VLUGR~\cite{blo1996}
The major conclusion from this failure is that the refinement takes
place at the wrong place, namely at the regions with steep gradients.
This is not where the dynamics of a pulled front
is determined. This occurs rather in the leading edge where any
perturbation of the linearly unstable state will grow \cite{ebe1997}.
Therefore the method in which the leading edge is included in 
the refined region gives even with a relatively low tolerance 
much better results than the standard refinement strategy.

In Table~\ref{velo} the front velocities $v_f$ are listed for the fine
($\Delta z=\frac{1}{4}$) and the coarse ($\Delta z=2$) uniform grids, as
well as for refined grids ($\Delta z^c=2$, $\Delta z^f=\frac{1}{4}$) 
with or without inclusion of the leading edge, 
and with linear or quadratic interpolation of the boundary conditions;
the tolerance is set to $\epsilon_\sigma=\epsilon_{\sigma-\rho}=10^{-3}$. 
The velocities in the table have all been determined from the displacement 
of the maximal electron density between $\tau$=250 and 262.5. 
As explained before~\cite{ebe1997}, the numerically computed front velocities 
should be smaller than the theoretical value,
which is indeed the case for the fine grids.

We also looked at the results obtained by discretizing the advective term
with a first-order upwind scheme. We concluded from those tests that 
this scheme performs very badly, quite in contrast to what is said 
in \cite{pan2003}. The amount of numerical diffusion introduced by that scheme
completely changes the asymptotic velocities on realistic grids.
Moreover, numerical diffusion can be expected to
over-stabilize the numerical scheme~\cite{hun2003} 
and to suppress thereby interesting
features of the solutions such as streamer branching.

\begin{table}[tb]
\caption{\footnotesize\sf Front velocities $v_f$ on various grids. 
The refinements have
been carried out with  $\epsilon_\sigma=\epsilon_{\sigma-\rho}=10^{-3}$. 
The exact asymptotic velocity is $v^\ast=1.3836$. 
For comparison, results for the 1st-order upwind scheme on the fine
grid are added.}
\label{velo}
\begin{center}
\small
\begin{tabular}{|l|c|}
\hline
\qquad\qquad\qquad method & $v_f$  \\ \hline \hline
uniform $\Delta z=\frac{1}{4}$, 1st-order upw. & 1.585  \\ \hline
uniform $\Delta z=\frac{1}{4}$, flux limited & 1.365  \\ \hline
uniform $\Delta z=2$,  flux limited & 1.448  \\ \hline
standard refinement  & 1.469 \\ \hline
refinement including leading edge & 1.365 \\ \hline
\end{tabular}
\end{center}
\end{table}

From these tests it appears that the grid refinement based on a simple
curvature monitor works well provided the leading edge is included in the
regions of refinement. This is in accordance with \cite{ebe1997} where the importance
of the leading edge for dynamics of a planar front is discussed.

\section{Performance of the code on streamer propagation with axial symmetry}
\label{sectest2d}
A planar ionization front as treated in the previous section is mainly of theoretical interest.
Genuine ionization fronts are curved around the streamer head, which leads to field enhancement 
ahead of the front.
We now consider the streamer propagation with cylindrical symmetry (2D case),
which differs substantially from the planar front (1D case), mainly due to the
field enhancement ahead of a curved front. In particular,
the electric field cannot be calculated as easily as in Eq.~(\ref{Eint}), but has to be computed through the
Poisson equation for the electric potential.
The tolerance $\epsilon_{\sigma-\rho}$ will play a non-negligible role in that case.
Also, as the electric field is enhanced immediately ahead of the ionization front 
and decreases further ahead, the leading edge region of maximal linear instability of the non-ionized state is bounded.

In this section the refinement algorithm as presented in Sect.\,\ref{refsect} will be applied 
to a streamer initiated by a Gaussian ionization seed situated on the axis of symmetry at the
cathode, $z=0$.
\be
\sigma(r,z,0)=\rho(r,z,0)=10^{-4}\exp\left(-\frac{r^2+z^2}{100}\right)\, .
\label{ic2}
\ee
The computational domain is $L_r$=1024 and $L_z$=2048, and the background electric field
is set to $|\E_b|=0.5$.
We use homogeneous Neumann boundary conditions at the cathode ($z=0$), which, together
with the ionization seed placed on the cathode, means that electrons flow into the system.
For other boundary conditions we refer to Sect.~\ref{subsecbc}.

\begin{figure}
\begin{center}
\includegraphics[width=14cm]{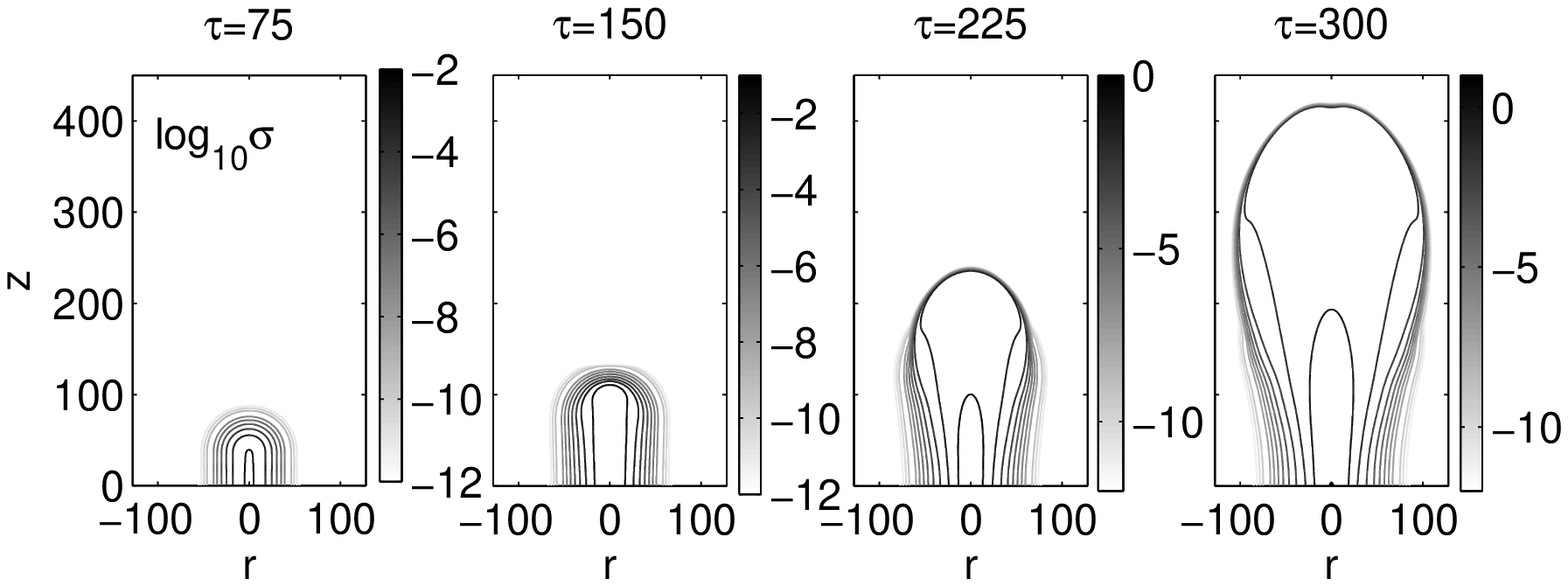}
\includegraphics[width=14cm]{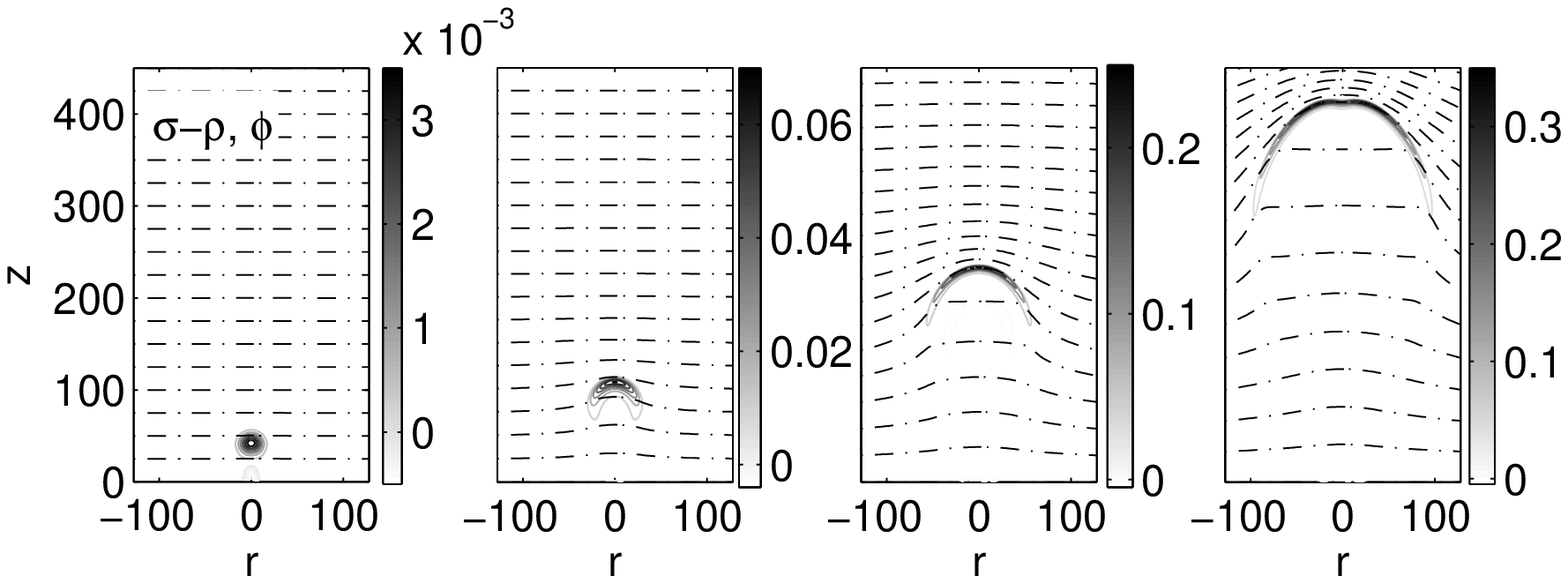}
\includegraphics[width=14cm]{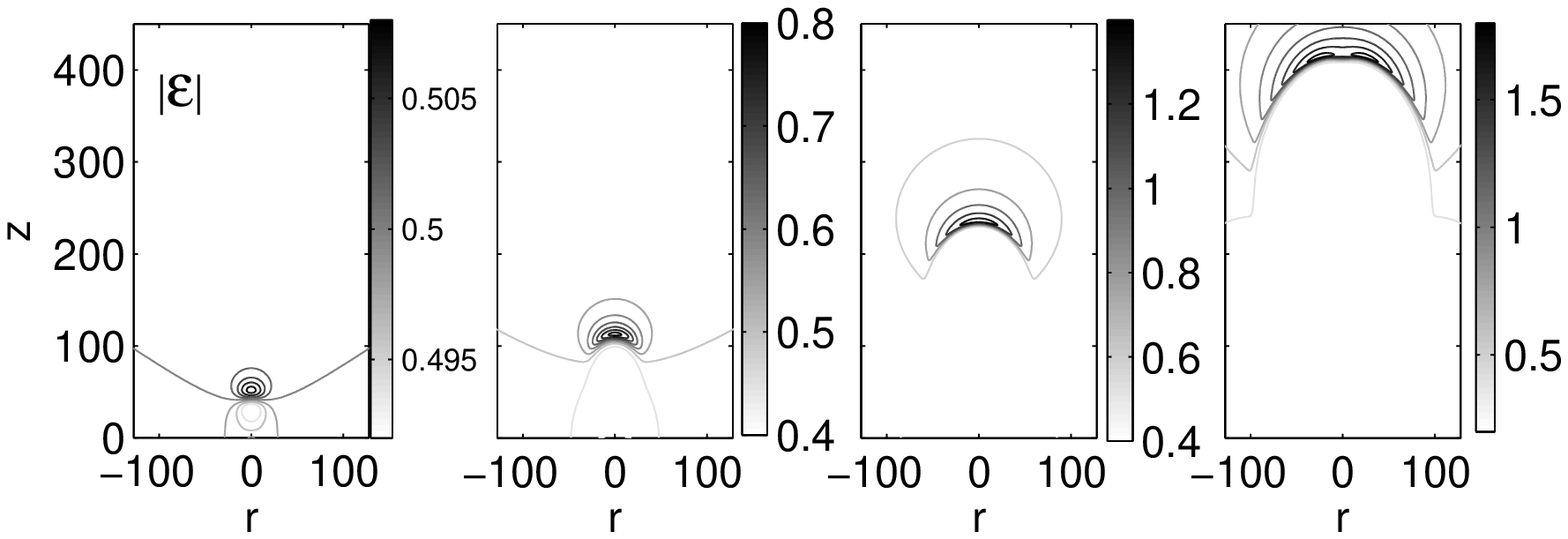}
\caption{\footnotesize\sf Propagation of a Gaussian ionization seed as given by Eq.(\ref{ic2}) in a 
background field $|\E_b|=0.5$. The snapshots correspond to times $\tau$=75, 150, 225 and 300.
The upper row gives the logarithm of the electron density $\sigma$, the middle 
row the net charge density $\sigma-\rho$ and the lower figure the electric field 
strength $|{\bs{\mc E}}|$. A uniform grid $\Delta r=\Delta z =1$ has been used for both the 
continuity equations and the Poisson equation.}
\label{streamerevol}
\end{center}
\end{figure}

The temporal evolution of the ionization seed~(\ref{ic2}) under these conditions is shown
in Fig.~\ref{streamerevol}. 
A uniform grid with grid size $\Delta r=\Delta z=1$ covers 
the domain where both the electron and positive ion densities are strictly positive. 
This is a relatively coarse grid, but, as mentioned 
in Sect.\,\ref{seclimuni}, the use of finer grids would require too many grid points 
for the {\sc Fishpak} routine to handle within an acceptable accuracy. Therefore we will use this 
uniform grid computation as a reference to test the performance of the adaptive refinement 
method. 
Fig.\,\ref{streamerevol} shows the snapshots of the electron and net charge density 
distributions as well as the electric field, at $\tau=75$, 150, 225 and 300. 

At $\tau=75$, the maximal deviation of the electric field strength 
from its background value is around 0.4\%, and space charges do not 
play a significant role yet. This is the electron avalanche phase, during 
which the Gaussian electron density distribution advects with an almost 
constant velocity, undergoes a diffusive widening, and grows due to impact 
ionization, leaving behind a small trail with positive ions~\cite{mon2005-1}. 

At $\tau=150$ the space charges concentrate in a layer around the streamer head, as 
can be seen in the second column of Fig.\,\ref{streamerevol}.  Due to its curvature, 
this space charge layer focuses the electric field towards
the axis of symmetry, thereby enhancing it. The body of the streamer is sufficiently ionized (and its
conductivity therefore enhanced) and the electric field in the streamer body is lower
than the background field. The space charge layer becomes thinner and denser in time,
and the electric field is increasingly enhanced, as can be seen in the third column
of Fig.\,\ref{streamerevol}. 

Eventually, the streamer becomes unstable, and branches. The beginning of this
branching state is shown in the rightmost column of  Fig.\,\ref{streamerevol}.

Let us now consider the effect of the refinements on the streamer dynamics, as well as the
effect of cutting off the densities that are below the continuum threshold. To this end, we
run the simulations with the same parameters -- background field, initial and boundary conditions --
as in the previous subsection, but we now allow one level of refinement for the continuity
equations, and seven levels of refinement for the Poisson equation. The finest grids in both cases
have a finest mesh size $\Delta r^f=\Delta z^f=1$. In the next section, more levels of refinement will be used.
In Sect.\,\ref{sec1D} we could see that, provided the leading edge of the streamer
front is included in the refinement, and a quadratic interpolation is used to provide the boundary conditions
for finer grids, a tolerance for the continuity equations of $\epsilon_\sigma=10^{-3}$ is well suited.
Moreover, the error in the spatial discretization of the net charge density $\sigma-\rho$ is 
induced by the discretization of the drift and diffusion terms in Eq.(\ref{sigmaeq}).
Hence it is natural to take the tolerance for the net charge density equal to that for the electron density.

The choice for the tolerance for the refinement of the Poisson equation is less straightforward. 
In one
dimension, the error in the second order discretization of the Poisson equation (\ref{phieq}) reads
\be
\delta\phi=\frac{\Delta z^2}{12}\frac{\partial^4\phi}{\partial z^4}=\frac{\Delta z^2}{12}\frac{\partial^2(\sigma-\rho)}{\partial z^2}\, .
\label{errphi1d}
\ee
Therefore, in the one-dimensional case, the curvature monitor for the net charge density will also
give the error in the spatial discretization of $\phi$. In higher dimensions however, this correspondence
does not strictly hold anymore. Nevertheless we assume that the tolerance for $\sigma-\rho$ will 
still give a good estimation for the error in the solution of the Poisson equation, and we therefore take
\be
\epsilon_\phi=\epsilon_{\sigma-\rho}=\epsilon_\sigma=10^{-3}\, .
\label{tol2d}
\ee
The number of grid points in the $r-$direction contained in each fine grid for the continuity equations 
has be chosen as $M_0=32$.
\begin{figure}[b]
\begin{center}
\includegraphics[width=13cm]{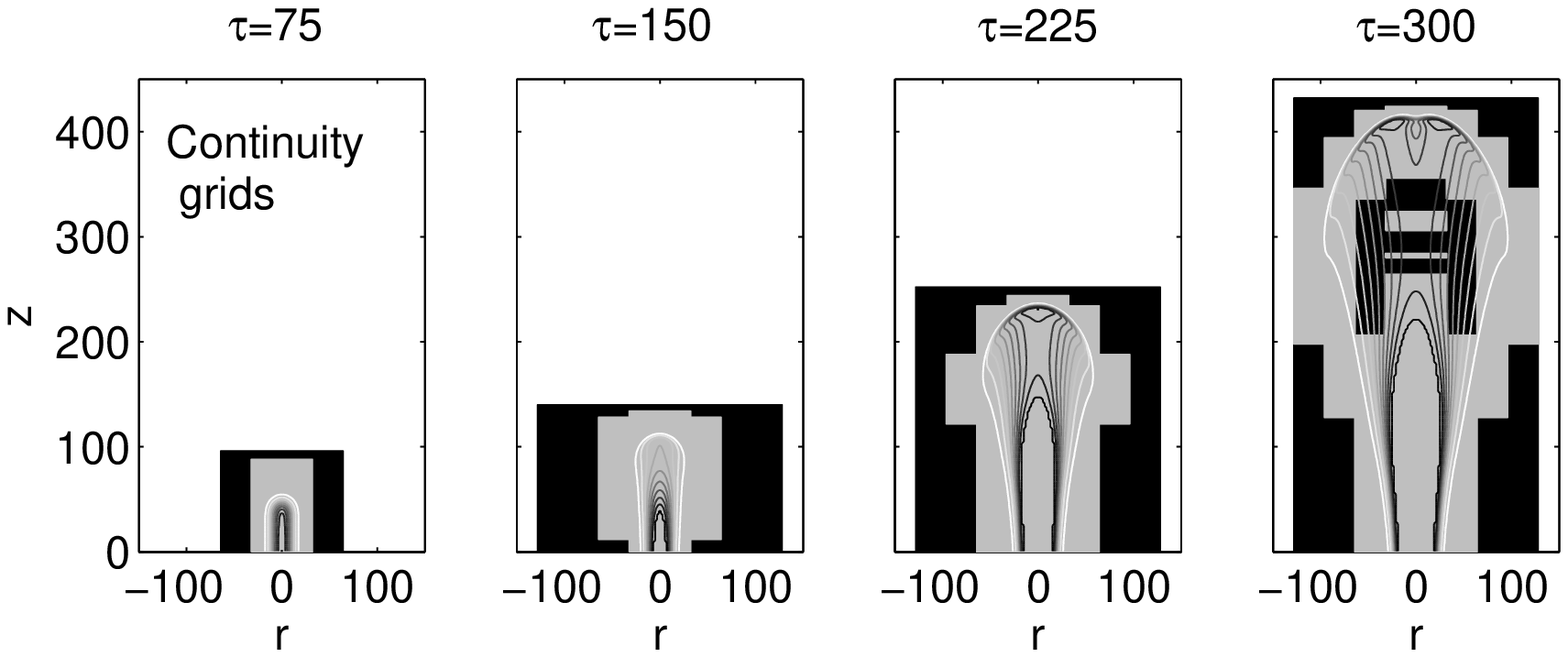}
\includegraphics[width=13cm]{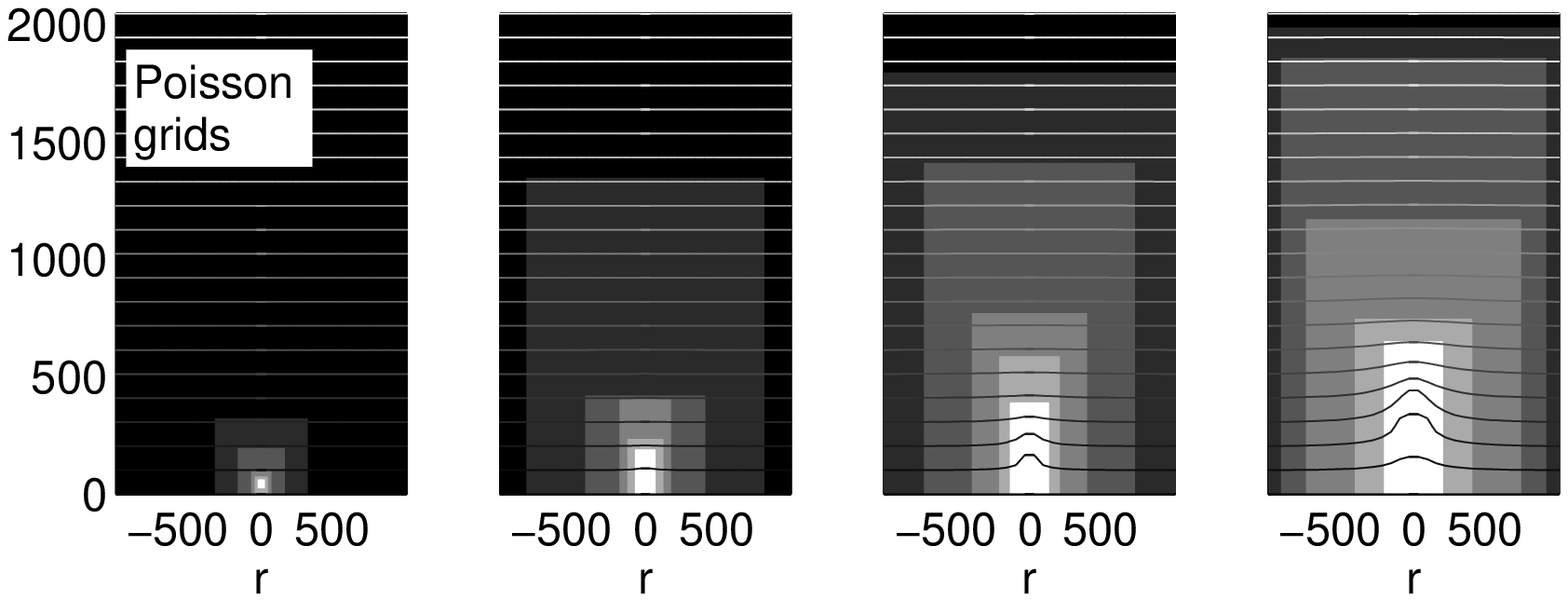}
\caption{\footnotesize\sf Grid distributions for the continuity equations (upper row), 
and for the Poisson equation (lower row).
The times are the same as in Fig.\,\ref{streamerevol}.
For the continuity grids, the black regions correspond to the coarsest grid 
with $\Delta r^c=\Delta z^c=2$, and which covers the domain on which the particle densities are
above the continuum threshold. The gray regions are covered with the fine grids with $\Delta r=\Delta z=1$.
The two coarsest grids -- with $\Delta r=\Delta z=128$ and 64 -- on which the Poisson equation has 
been solved cover the whole computational domain, which is filled in black. The finer grids 
with cell size 32, 16, 8, 4 and 2 are plotted in lighter gray shades, the white grid corresponds 
to $\Delta r^f=\Delta z^f=1$. }

\label{gridscont}
\end{center}
\end{figure}
The net charge density distribution at $\tau=75$, 150, 225 and 300 are plotted in 
the upper row of Fig.\,\ref{gridscont},
together with the grids on which the continuity equations have been solved. The equipotential
lines are shown together with the grid distribution for the Poisson equation in the lower panel 
 of Fig.\,\ref{gridscont}.
\begin{figure}
\begin{center}
\includegraphics[width=15cm]{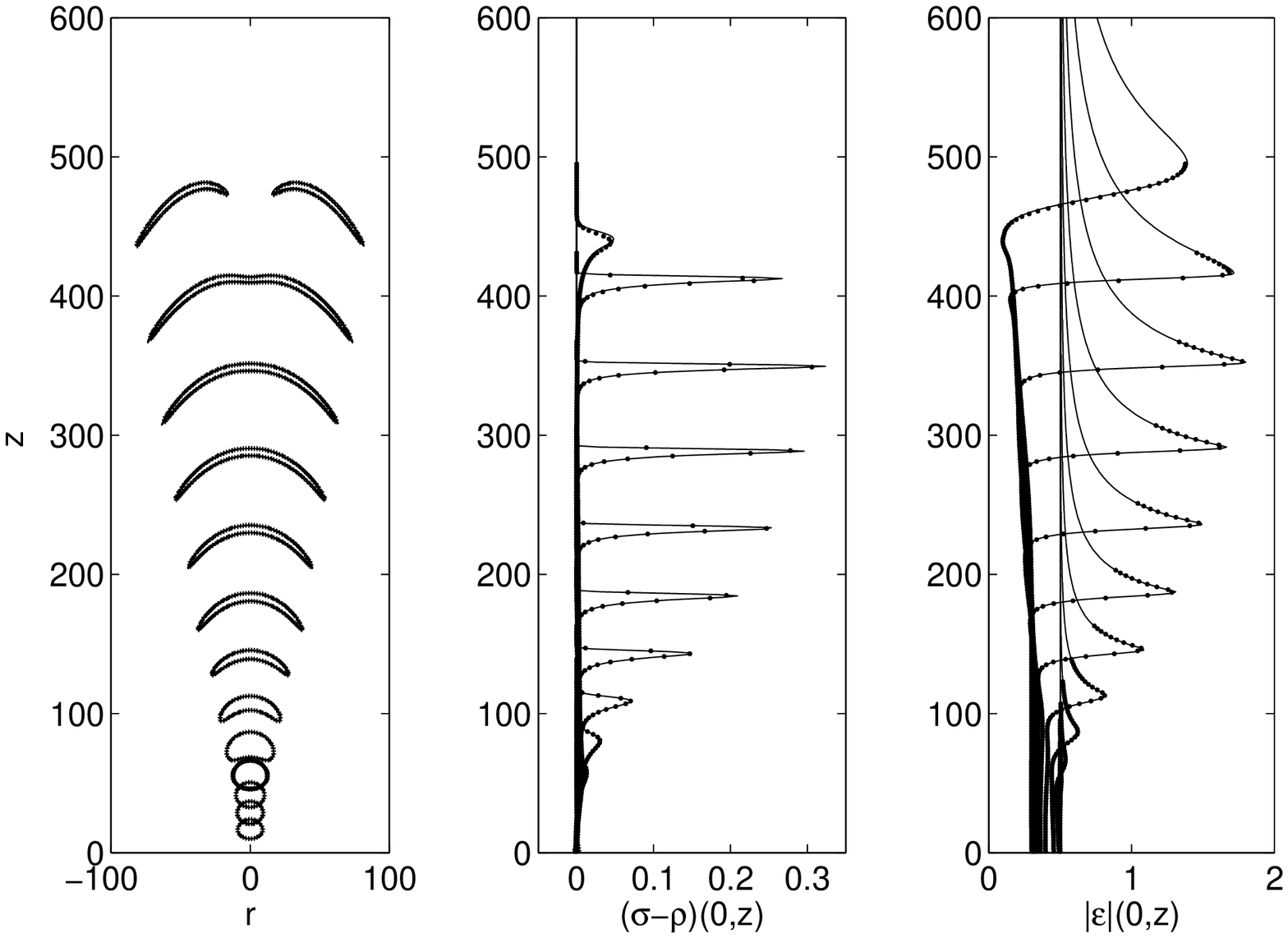}
\caption{\footnotesize\sf Comparison of uniform grid results with those obtained using the grid refinement strategy. 
Left panel: evolution of the half maximum contours of the net charge density.
Middle panel: evolution of the net charge density on the axis of symmetry, $(\sigma-\rho)(0,z,\tau)$.
Right panel: evolution of the electric field strength on the axis of symmetry, $|\E|(0,z,\tau)$.
The results in all panels are shown for  $\tau=25$ to 325, with steps of 25.
Solid line: uniform grid ($\Delta r=\Delta z=1$), dots: grid refinement ($\Delta r=\Delta z=1,2$).} 
\label{comp_charge2}
\end{center}
\end{figure}

Since the streamer front is not planar anymore, it is necessary to investigate the 
effect of the refinements not only on the axis, but in the whole streamer. 
The evolution of the half maximum contours of the net charge density is depicted in 
the left plot of Fig.\,\ref{comp_charge2}, both in the uniform grid computation without 
cut-off below the continuum threshold, and with
the refinements including the cut-off. 
There is an excellent correspondence between the uniform grid computation and the 
one where all grids are refined. At some places there is a slight difference between the results,
but this is also a consequence of the results being plotted for the coarsest grid, i.e.
$\Delta r^c=\Delta z^c=1$ for uniform grid, and on $\Delta r^c=\Delta z^c=2$ for the refined grids. The 
grid points on which the results are plotted therefore do not coincide, and the contours consequently 
might be slightly different. Up to branching however, there is no significant error in either the 
radius or the width of the space charge layer. A minor effect of the refinement only becomes
visible once the streamer has become unstable.  A more detailed
investigation of this branching will follow in a later paper.

The leftmost plot of Fig.~\ref{comp_charge2} has to be completed with the evolution of 
the axial charge density distribution of the 
streamer, in order to control not only the position of the half maximum contours, but
also the maxima themselves. This is shown in the middle and rightmost plots of 
Fig.\,\ref{comp_charge2}, where the axial
charge density and electric field strength distributions, respectively,
 are shown for the same times as the leftmost plot of Fig.\,\ref{comp_charge2}.
Again it appears that the adaptive grid refinement method produces 
results that coincide very well with the uniform grid computation.

\begin{figure}
\begin{center}
\includegraphics[width=10cm]{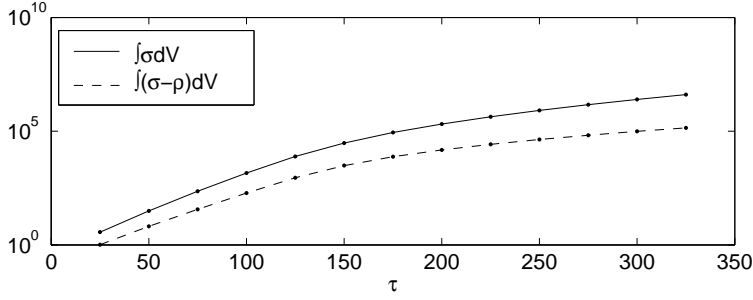}
\caption{\footnotesize\sf Evolution of the total number of electrons (solid line) and net charge (dashed line). The lines
correspond to a uniform grid computation without cutting off the densities below the continuum threshold, the dots
to the refined grid computation.}
\label{comp_N}
\end{center}
\end{figure}

Finally, because of the charge conservation during an ionization event, it should be verified that 
the refinements are indeed mass conserving, as has been taken care of in Sect.\,\ref{secrefcont}. This is 
shown in Fig.\ref{comp_N}, where we have compared a uniform grid computation where the densities 
below the continuum threshold have not been cut off, with the refined grid computation. It is clear that
neither the cut-off, nor the refinement disturb the total number of particles in a visible way.

\begin{figure}
\begin{center}
\includegraphics[width=10cm]{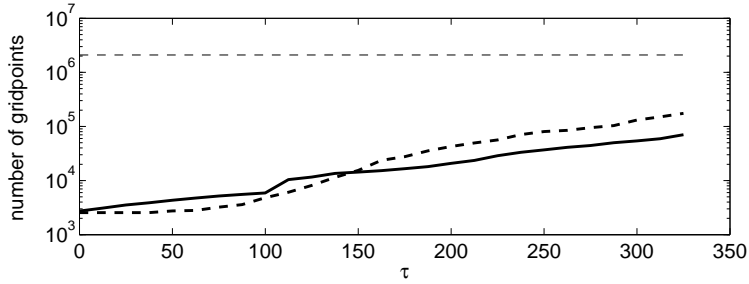}
\caption{\footnotesize\sf Number of grid points as a function of time. The thin dashed line represents the number of grid points
corresponding to a grid of grid size $\Delta r=\Delta z=1$ over the whole domain, i.e. 1024$\times$2048.
The thick lines give the temporal evolution of the number of grid points when the refinement algorithm
is applied up to a finest mesh size of $\Delta r=\Delta z=1$. {\em Solid line}: continuity equations,
1 level of refinement. {\em Dashed line}: Poisson equation, 6 levels of refinement.}
\label{countpoints}
\end{center}
\end{figure}

The gain in memory obtained with the refinement method can be observed through the number of grid points that
are used to solve the continuity equations and the Poisson equation. As can be seen in Fig.\ref{countpoints},
the number of grid points used are one to three orders of magnitude smaller than in the computations
as performed in \cite{roc2002}, where a uniform grid covers the whole computational domain. For the
gain in CPU time, we refer to table~\ref{CPU} in the next section.

We emphasize that in the previous test the choice for two levels of refinements has been made 
in order to compare the results with uniform grid computations. In later simulations we will
use more refinement levels, thereby reaching much smaller mesh sizes. We can however extrapolate the
outcome of these tests to cases using more levels.
\section{Accuracy requirements for the streamer simulations}
\label{secacc}

To illustrate the use of the above algorithm, we present results
in a new parameter regime, namely in a very long gap with relatively
low background field, and a short gap with high field.
We present results on different finest mesh sizes, and investigate the convergence of
the solution on decreasing the finest mesh.

\subsection{Long streamers in a low electric field}
                                                                                                    
We consider a gas gap on which a background electric field $|\E_b|=0.15$ is applied.
The negative electrode (cathode) is placed at $z=0$, the positive one (the anode)
at a distance $L_z=2^{16}=$65~536. The radial boundary is situated at $L_r=2^(15)=$32~768.
For N$_2$ at atmospheric
pressure, this corresponds approximately to an inter-electrode distance of 15 cm
and an electric field of 30 kV/cm.

The initial seed~(\ref{ic}) is placed on the cathode at $z=0$ ($z_b=0$). The maximal initial density is
$\sigma_0=1/4.7$ and the e-folding radius of the seed is $R_b=10$, which correspond to
a density of 14 cm$^{-3}$ and a radius of 23 $\mu$m, respectively, for N$_2$ under normal conditions.
This relatively dense seed enables us to bypass the avalanche phase. If we
would start with a single electron at this value of the electric field, analytical results~\cite{mon2005-1}
show that the transition to streamer would occur after the
avalanche has traveled a distance $d\approx18\alpha^{-1}(|\E_b|)\approx$ 14~000.
Using a dense seed accelerates the transition time considerably. The dense seed mimics streamer
emergence from a needle inserted into the cathode or from a laser induced ionization seed.

The coarsest grid for the continuity equations has a mesh size of 64 in both directions, the coarsest
one for the Poisson equation a size of 8192.
We first present the results when the grids are refined up to a mesh size $\Delta r^f=\Delta z^f=2$.
\begin{figure}
\begin{center}
\includegraphics[width=14cm]{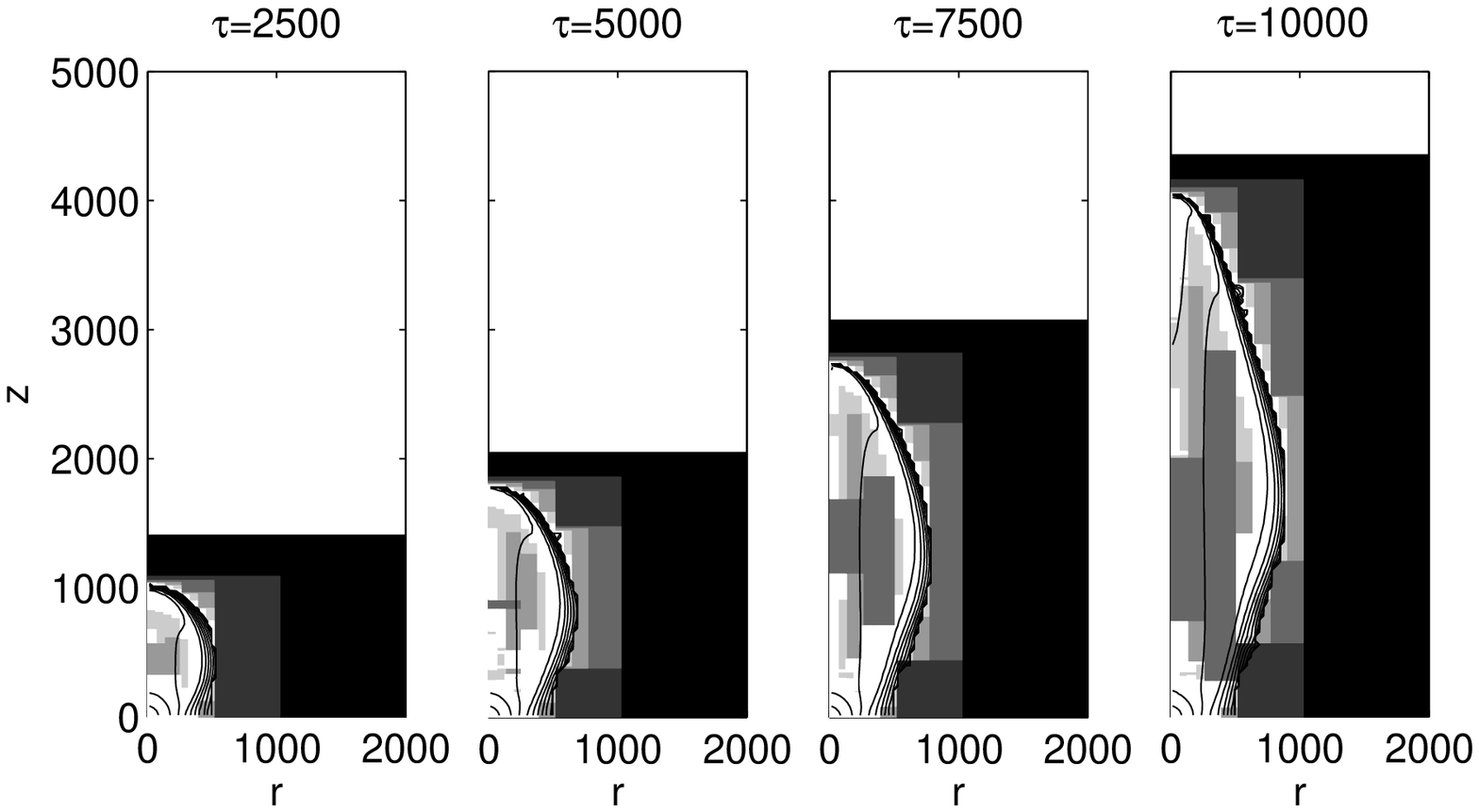}
\includegraphics[width=14cm]{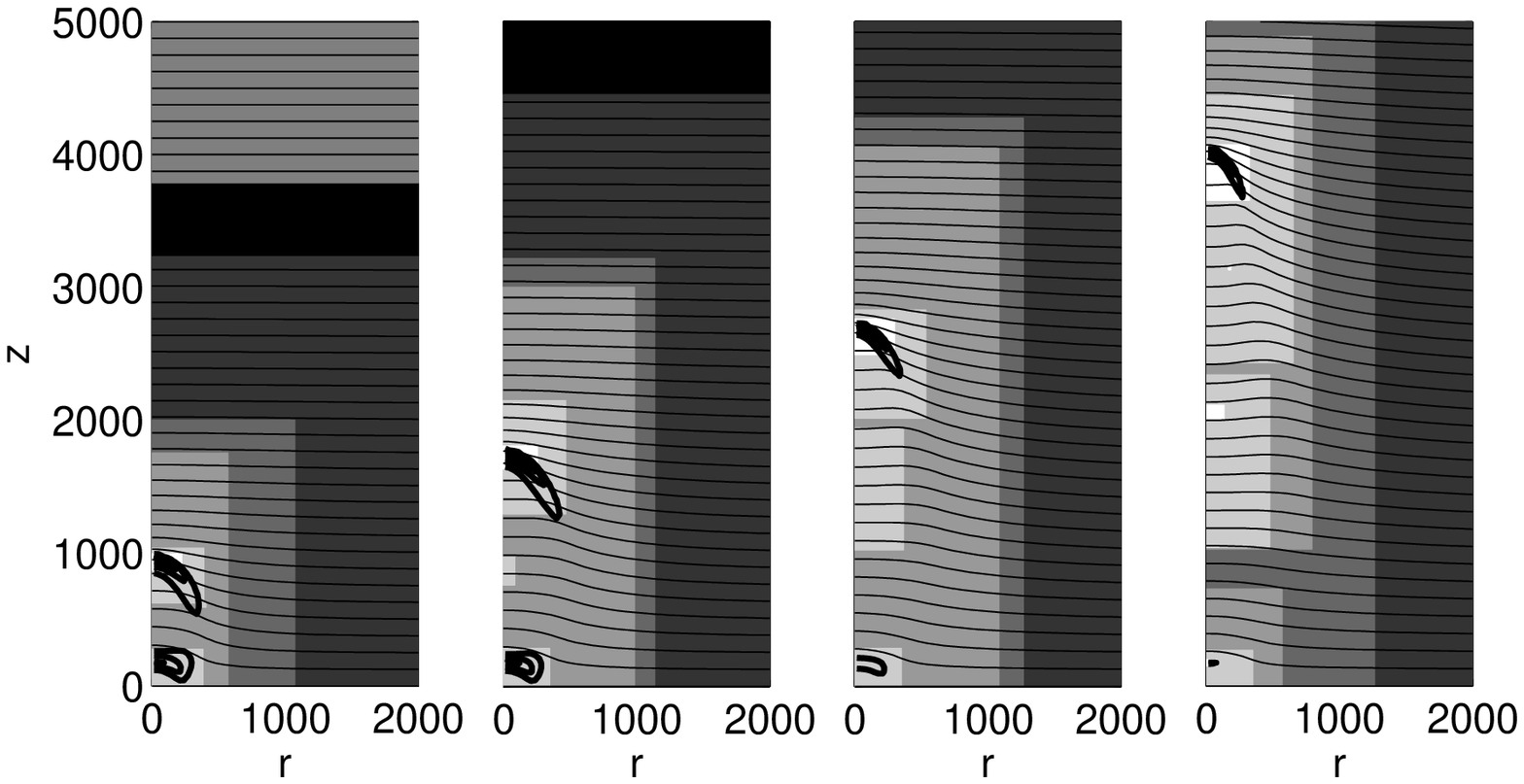}
\caption{\footnotesize\sf Evolution of a streamer and the computational grids in a low background electric field
$|\E_b|=0.15$. Upper panel: logarithm of the electron density with the grids for the 
continuity equations. The coarsest grid with mesh size 64, covering the domain on which 
the continuum approximation for the densities holds, is filled in black. The finest grids with
mesh size 2 are white. The particle densities in the white region ahead 
of the streamer are below the continuum threshold and therefore not solved.
Lower panel: net charge density (thick lines)
and equipotential lines (thin lines) together with the grids for the Poisson equation. The white
domains are covered with the finest mesh size of 2, the black regions are covered by
a grid with mesh size 64. The gray region ahead of the streamer at $\tau$=2500 is covered by
a grid with mesh size 128. The coarser grids (with mesh sizes up to 8192) are not shown.
}
\label{evolE015ic100}
\end{center}
\end{figure}
                                                                                                    
The upper panel of Fig.\,\ref{evolE015ic100} shows the electron density distribution
on a logarithmic scale. There are two regions in the streamer: a rather wide one with
low electron densities, and one which is much narrower and very dense.
The lower panel of Fig.\,\ref{evolE015ic100}
clearly shows the negatively charged layer and its effect on the equipotential lines. These
are close to each other ahead of the streamer tip, which indicates an enhancement of the electric field.
In the interior field the distance between the equipotential lines is slightly larger than
outside the streamer, which implies that the field in the streamer body is somewhat
smaller than the background field.
                                                                                                    
Let us now compare these results with those obtained on coarser grids. We have run the simulations
on grids that were refined up to mesh sizes of $\Delta r^f=\Delta z^f=4$ and $\Delta r^f=\Delta z^f=8$, 
thus the finest grids in these cases
are twice respectively four times coarser than those on which 
the results shown in Fig.\,\ref{evolE015ic100}
have been obtained.
                                                                                                    
\begin{figure}
\begin{center}
\includegraphics[width=15cm]{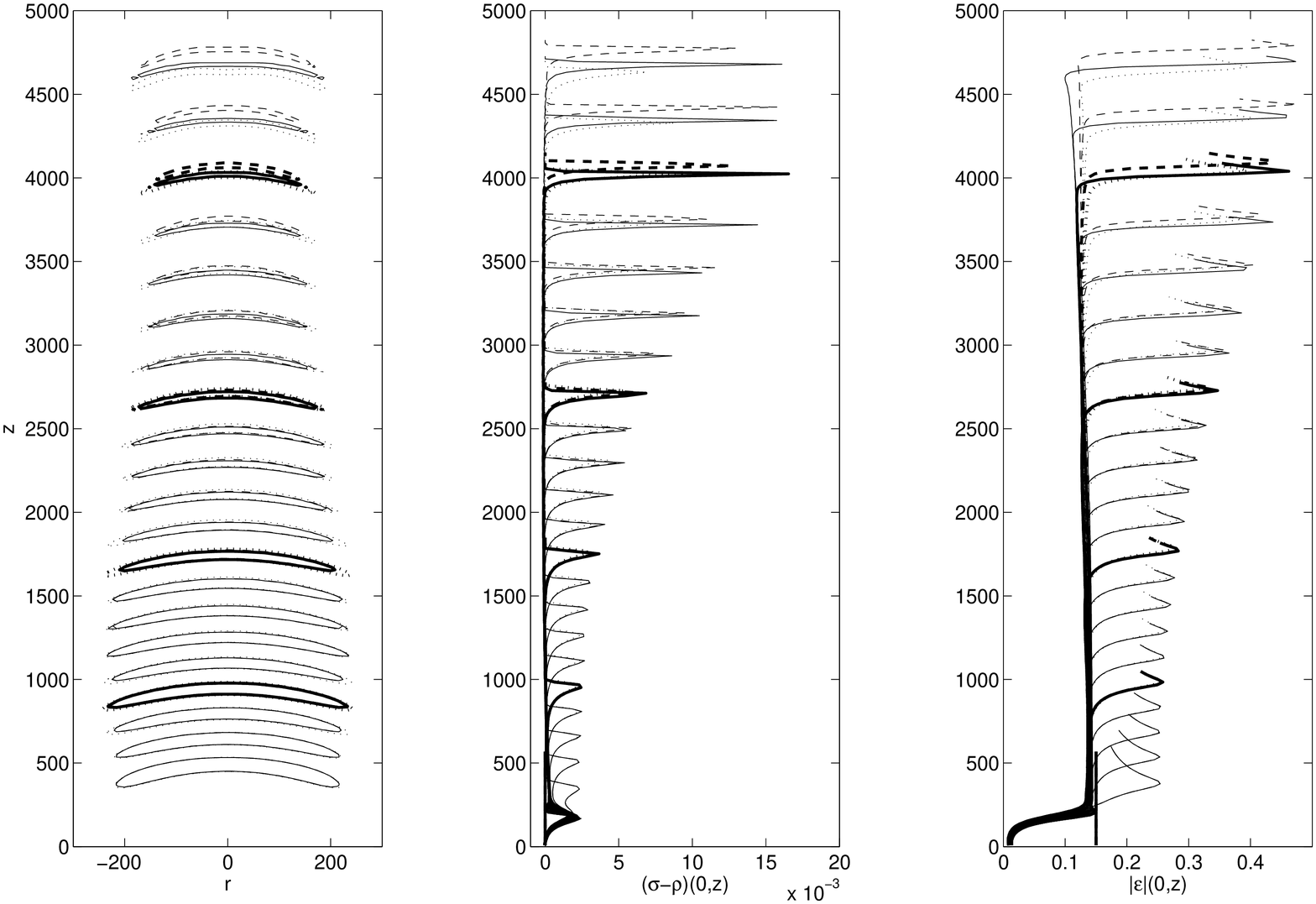}
\caption{\footnotesize\sf Influence on the mesh size on the numerical solution for a background field of 0.15.
The left panel shows the evolution of the half maximum contours of the net charge density $\sigma-\rho$. 
To show the structure clearly, the aspect ratio is not equal.
The middle and the right panels show the evolution of the net charge density and the electric field
strength on the axis, respectively. The times shown go from 500 to 11000 with equidistant  time steps
of 500. The thick lines correspond to the times shown in Fig.~\ref{evolE015ic100}. 
Solid line: $\Delta z^f$=2; dashed line: $\Delta z^f$=4; dotted line: $\Delta z^f$=8.}
\label{compa015}
\end{center}
\end{figure}

Fig.\,\ref{compa015} shows the influence of the mesh size by means of the net charge density
distribution and the electric field. The leftmost plot of this figure shows the
evolution of the half maximum contours of the net charge density. The evolution of
the density distribution and the electric field strength are shown in the middle and rightmost plot, 
respectively. Up to $\tau\approx5000$ we see that the coarse grid simulations ($\Delta z^f=8$) already 
give convergent results. After this time, the front starts to move somewhat too fast
on this coarse grid. This is due to numerical diffusion. The simulation with $\Delta z^f=4$ 
gives the same results as those on a grid with size $\Delta z^f=2$ up to $\tau\approx 7500$, after which
numerical diffusion again makes the front move somewhat too fast. 
We emphasize However, the influence
of the grid becomes significant only around $\tau\approx10000$, when
the streamer head becomes unstable. The effect of the  grid size  on the streamer instability
will be considered in more detail in a future paper.

We notice that the solution on a grid with mesh spacing 8 eventually lags behind the fine grid 
computation. This non-monotonous
behavior is due to a competition between two opposite effects: the numerical diffusion tends
to accelerate the field, but also to reduce the maximum of the charge distribution, thereby 
reducing the maximal electric field strength, and thus the propagation velocity of the streamer.

\begin{figure}
\begin{center}
\includegraphics[width=10cm]{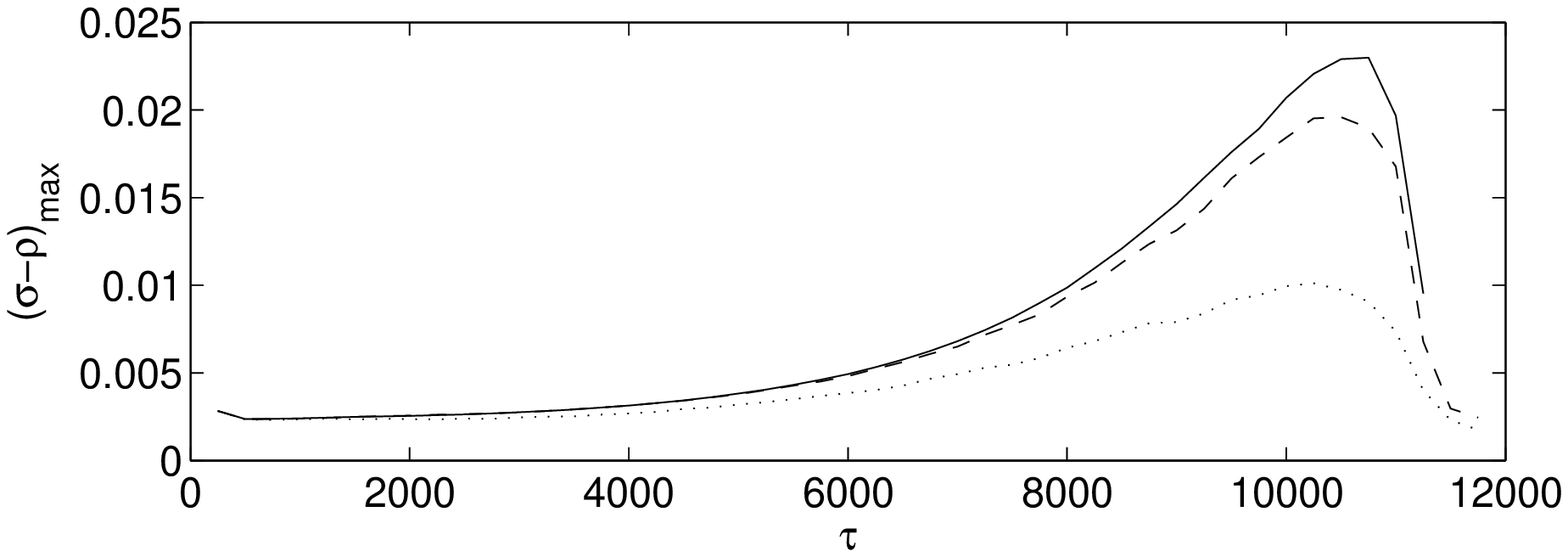}
\caption{\footnotesize\sf Temporal evolution of the maximal net charge density on the axis. 
Solid line: $\Delta z^f$=2; dashed line: $\Delta z^f$=4; dotted line: $\Delta z^f$=8.}
\label{compdum}
\end{center}
\end{figure}

In Fig.~\ref{compdum} we compare the evolution of the maximal net charge density on the 
axis of symmetry for different choices of the finest mesh size. It shows that, the coarser the grid,
the more underestimated the maximal net charge density becomes. This
indicates that it is indeed numerical diffusion which smears out the space charge layer,
thereby accelerating it and decreasing its density.

We conclude that the simulations run on a finest grid with mesh size 8 will give non-negligible
errors in the results. However, one more level of refinement, leading to a finest grid
with mesh size 4, already gives acceptable solutions during the streamer regime.
Moreover, from these results, we can extrapolate that refining the grids even more (e.g.
up to a finest mesh size of 1) will not lead to a significant correction of the results, and
that the computational work required to run the simulations on such fine grids is not
in proportion with the gain in accuracy.  

\subsection{Fine grid computations at a high electric field}
\label{sectesthighfield}
We now consider the evolution of a streamer in a short overvolted gap.
The inter-electrode distance is set to 2048, corresponding to 5~mm
for N$_2$ at atmospheric pressure. The background electric field is set to $\E_b=-0.5{\bf\hat e}_z$, which corresponds to 100 kV/cm.
The initial seed~(\ref{ic}) is placed on the cathode ($z_b=0$). 
The maximal density of the initial seed is set to $\sigma_0=10^{-4}$, and its 
radius to $R_b=10$.

\begin{figure}
\begin{center}
\includegraphics[width=13cm]{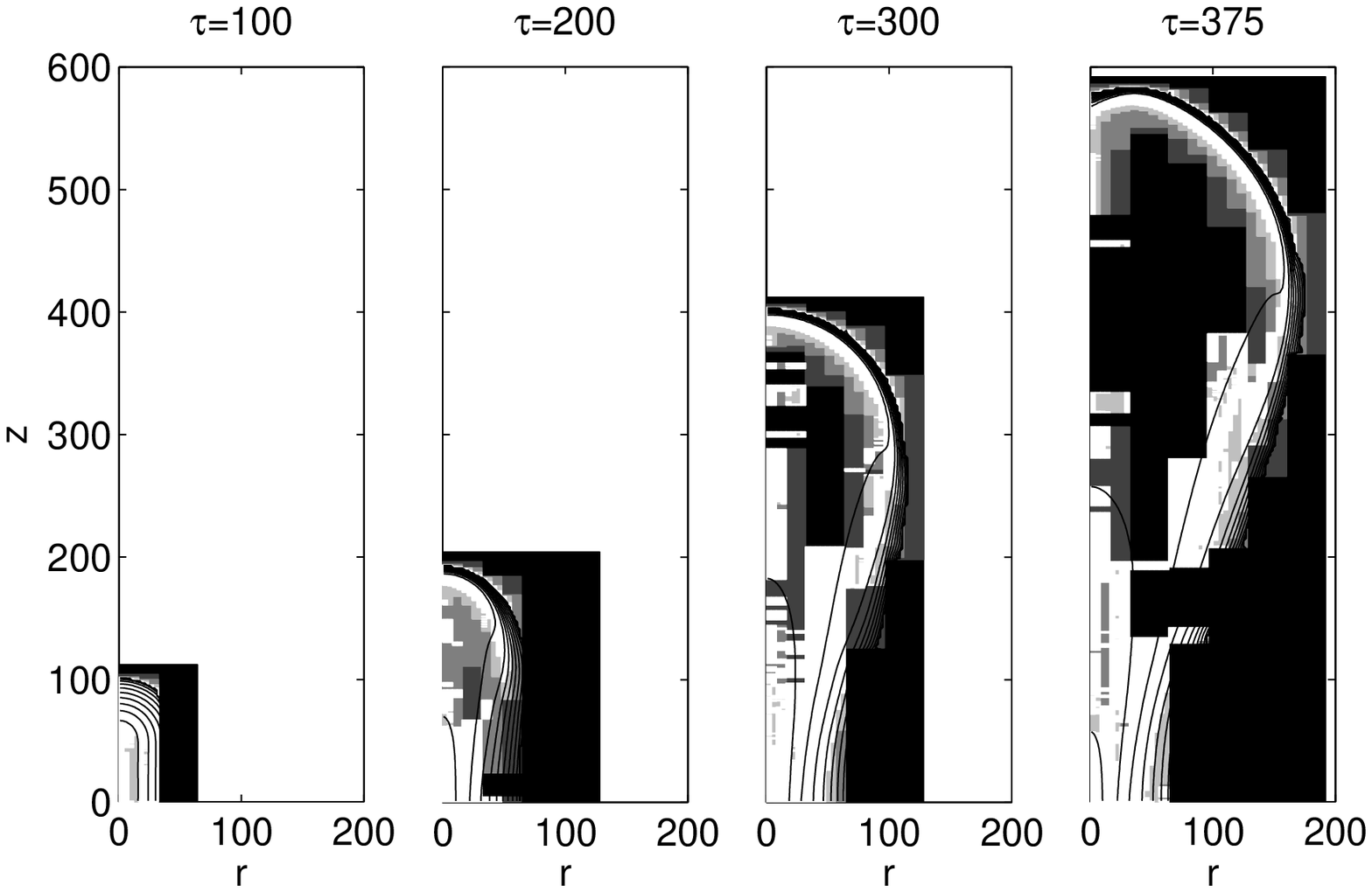}
\includegraphics[width=13cm]{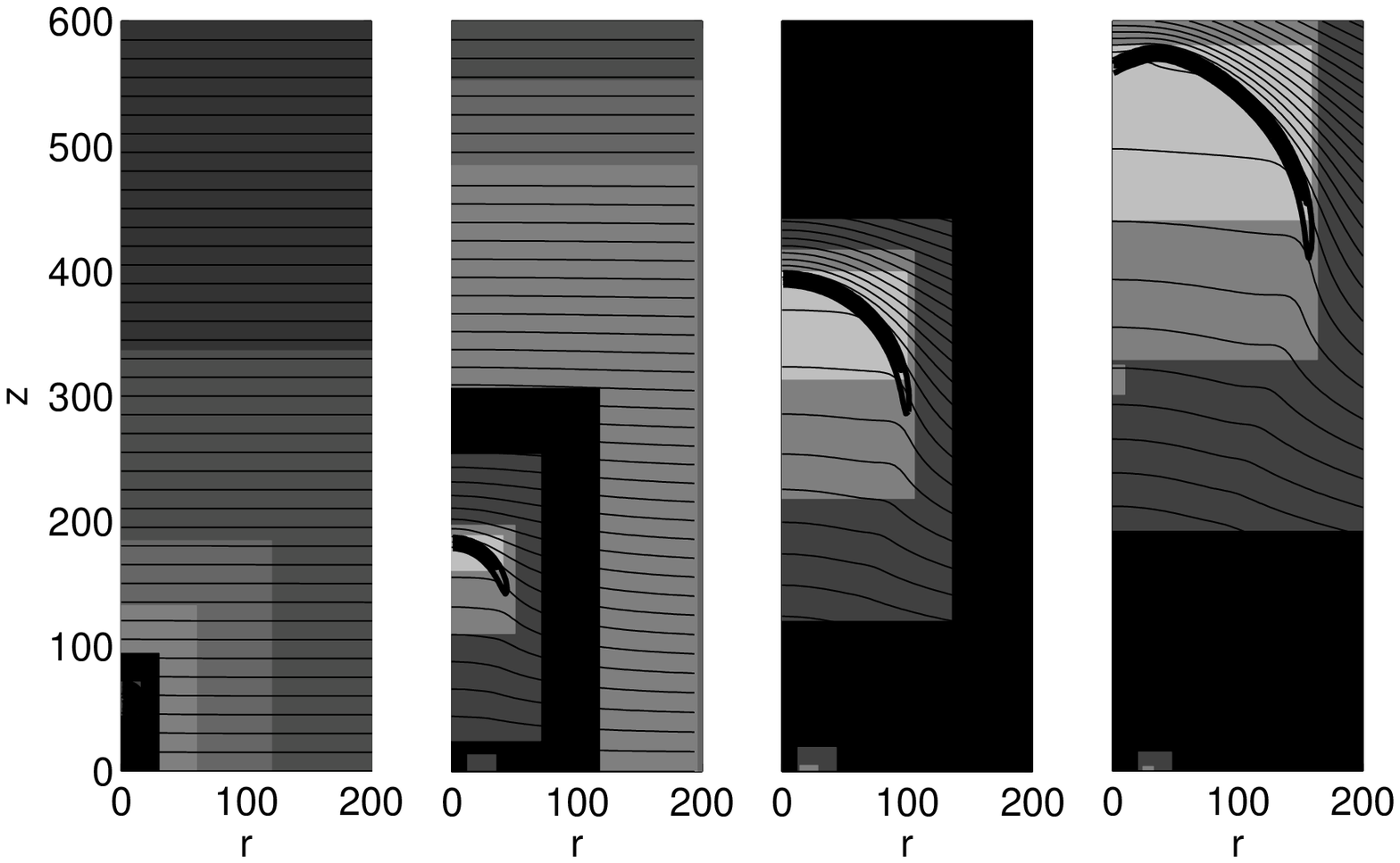}
\caption{\footnotesize\sf Evolution of a streamer and the computational grids in a high background electric field
$|\E_b|=0.5$. Upper panel: logarithm of the electron density together with the grids for the 
continuity equations. The coarsest grid has a mesh size $\Delta z^c=2$ and is filled in black. The finest
grid has a mesh size of $\Delta z^f=1/8$ and is white. The white region surrounding the coarsest grid has not
been covered by a computational grid since the densities are below
the continuum threshold there. Lower panel: net charge density (thick lines) and 
equipotential lines (thin lines) together with the grids for the Poisson equation. 
The black region is covered with a grid with mesh size 2, coarser grids are not shown on this scale. 
The grids are refined up to a mesh
of $\Delta z^f=1/8$. However, this finest grid is only used at $\tau$=375, in a very small part 
of the computational domain. The finest grid for the Poisson equation at $\tau$=100 has a mesh size of 1.
At $\tau$=200 and 300 the finest grid has a mesh size of 1/4.}
\label{evolE05hf0125}
\end{center}
\end{figure}
The evolution of the streamer with previous initial and boundary conditions,
 during the non-linear phase is shown in Fig.\,\ref{evolE05hf0125}. These results
have been obtained on finest grids with a mesh size of $\Delta z^f=1/8$ for both the
continuity and the Poisson equations. The coarsest grid for the continuity equation
has a mesh size of $\Delta z^c=2$, the one for the Poisson equation has a size of $\Delta z^c=128$.
                                                                                                    
The discharge clearly exhibits the streamer features which are, as in the
low field case, a thin negatively charged layer that enhances the electric field ahead of
it, and  partially screens the interior of the streamer body from the background electric field.
                                                                                                    
\begin{figure}
\begin{center}
\includegraphics[width=14cm]{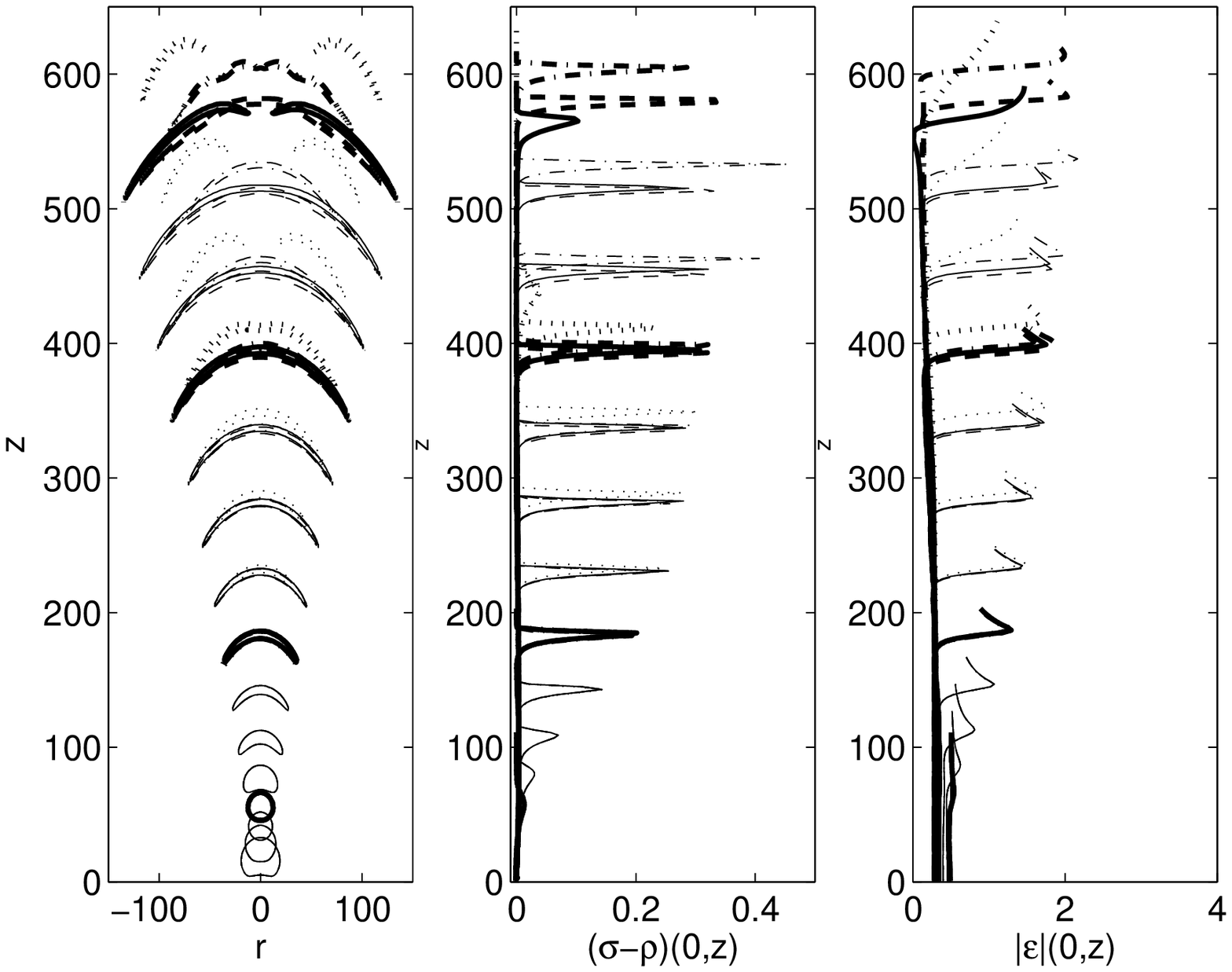}
\caption{\footnotesize\sf Influence on the mesh size on the numerical solution for a background field of 0.15.
The left panel shows the evolution of the half maximum contours of the net charge density $\sigma-\rho$.
The middle and the right panels show the evolution of the net charge density and the electric field
strength on the axis, respectively. The times shown go from 25 to 375 with equidistant time steps
of 25. The thick lines correspond to the times shown in Fig.~\ref{evolE05hf0125}.
Solid line: $\Delta z^f$=1/8; dashed line: $\Delta z^f$=1/4; dash-dotted line: $\Delta z^f$=1/2, dotted line:  $\Delta z^f$=1.}
\label{comp_vf}
\end{center}
\end{figure}

In order to investigate the convergence of the solution under decreasing 
grid size, we have also run the simulations on finest mesh sizes 
$\Delta z^f=1/4$, $\Delta z^f=1/2$ and $\Delta z^f=1$.
Fig.\,\ref{comp_vf} shows the evolution of the half maximum
contours of the net charge density, as well as that of the axial distribution of the net
charge density and the electric field strength.
                                                                                                    
We see that up to $\tau=200$, a grid size $\Delta z^f=1$ gives the same results
as a grid size $\Delta z^f=1/8$. After that time, the predicted front velocity on
a grid size of 1 is much faster than that on the finer grids, whereas the maximal electric
field does not differ that strongly from fine grid computations. This implies that the
front propagates faster because of numerical diffusion. 

The finer grids on the other hand all give similar solutions up to
$\tau\approx300$. After this time, the streamer head has become unstable, and the
numerical details will influence the branching behavior of the streamer. Indeed,
for $\Delta z^f=1/8$, the streamer exhibits two branches that propagate off-axis, whereas
for $\Delta z^f=1/4$ and 1/2 one branch continues propagating along the axis while the other does not.
Therefore the results at the time of branching differ. 
An off-axis branching results in a decrease of the maximal field and densities on the axis,
whereas an on-axis branching on the contrary results in an increase of these quantities.
This is why the maximal net charge density and field strength on axis are smaller in
the  $\Delta z^f=1/8$ case than in the  $\Delta z^f=1/4$ and 1/2  cases.

We emphasize that the different branching form is due to the instable nature of the streamer:
the streamer approaches a bifurcation point, and the further evolution is set by minor details,
like the numerical grid, in a chaotic manner. There is, however, convergence of the
branching time, which shows that unlike the behavior after branching, the onset of the instability
is not triggered numerically. 

In Table~\ref{CPU} we summarize the efficiency and accuracy of the refinement algorithm in terms 
of the CPU time, memory usage and accuracy in the spatial electron distribution 
for different finest mesh spacings and refinement levels. To this end we start the simulations from a 
well developed streamer and run the different cases over a time $T=10$, with a time step 
$\Delta\tau=5\cdot10^{-2}$. This time step is small enough to get negligible temporal errors. The initial 
particle distribution is the numerical solution at $\tau=200$ obtained on a hierarchy of grids 
with a finest and coarsest mesh widths $\Delta z^f$=1/8 and $\Delta z^c$=2 for the continuity equations
(see Fig.~\ref{evolE05hf0125}). For the Poisson equation, the coarsest mesh width 
was set to 64, and the finest on to 1/8. The solution at $\tau=210$ with these grid 
settings is taken as the reference solution for the computation of the errors.  
This benchmark has been performed on the node of linux cluster, a 32 bit Opteron 1.4 Ghz 
with 16 Gb memory. 

The cases with refinements all use the same coarsest mesh spacings and refinement tolerance, 
the finest mesh spacing varying from $\Delta z^f$ = 1/8 to $\Delta z^f$ = 1. 
To illustrate the performance of the algorithm, we show results 
on  uniform grids with spacing 1 and 2. A solution on a uniform grid of spacing 1/2 could
not be obtained because the Poisson solver would become very inaccurate (see Appendix A). 
We further note than on a regular pc (rather than the cluster node with the large
amount of memory used for the benchmark) memory would also become a limiting factor.

Since the grid configuration, and therefore the CPU time used per time step, is not constant,
the average values over the run of the number of  grid points and the CPU time are shown. 
The accuracy of the method can be characterized by the discrete
$L_1$, $L_2$ and $L_\infty$-norms of the errors. For a grid function
$e=(e_{ij})$ on a $m_r\times m_z$ grid these norms are defined as
\be
\begin{array}{l}
\displaystyle
||e||_1 =  \sum_{i,j}\Delta r\Delta z|e_{i,j}| \, , \quad 
\displaystyle
||e||_2 = \sqrt{\sum_{i,j}\Delta r\Delta z|e_{i,j}|^2}\, ,\quad 
\displaystyle
||e||_\infty = \max_{i,j}(|e_{i,j}|)\, .
\end{array}
\label{errL}
\ee
The errors are computed on the coarsest grid level $\Delta z^c=2$, after having restricted 
the values from finer grids where needed using Eq.~\ref{restr}.

\begin{table}[t]
\caption{\footnotesize\sf Performance of the refinement algorithm in terms of CPU time, memory
usage and errors in the solution. Shown are results on refined grids (four upper rows),
and on uniform grids (three lower rows). $\Delta z^f$ is the finest allowed mesh spacing for
both the continuity and the Poisson equations. $n_{lev}$ is the number of levels for the
continuity equation, $n_{lev}=1$ corresponding to a uniform grid computation, for both Poisson
and continuity equations. $N_\sigma$ and $N_\phi$ are the number of grid points used for 
the continuity and the Poisson equations, respectively. The reference solution for the errors 
computation is the one obtained with $\Delta z^f$=1/8. An $\ast$ denotes a case which could 
not be tested because of the inaccuracy of the Poisson solver and/or the lack of computational memory.}
\label{CPU}
\vspace{1.4mm}
\begin{center}
\begin{tabular}{|c|c|c|r|r|c|c|c|c|}
\hline
$\Delta z^f$ & $n_lev$ & CPU time (s)& $N_\sigma$    & $N_\phi$    & $||e||_{L_1}$ & $||e||_{L_2}$ & $||e||_{L_\infty}$ \\
\hline
1/8          & 5  & 11.06  & 657 856 & 93 584 &                      &                    &                     \\
\hline
1/4          & 4  & 9.68   & 193 024 & 93 584 & $5.09\cdot10^{-9}$ & $1.18\cdot10^{-6}$ & $6.13\cdot10^{-4}$ \\
\hline
1/2          & 3  & 6.67   & 76 160 & 69 452 & $2.71\cdot10^{-8}$ & $7.41\cdot10^{-6}$ & $4.62\cdot10^{-3}$ \\
\hline
1            & 2  & 4.58   & 21 632 & 44 248 & $1.04\cdot10^{-7}$ & $2.78\cdot10^{-5}$ & $1.64\cdot10^{-2}$ \\
\hline
1/2          & 1  & $\ast$ & $\ast$          & $\ast$          & $\ast$             & $\ast$             & $\ast$             \\
\hline
1            & 1  & 7.24   & 2 097 152 & 2 097 152 & $1.54\cdot10^{-7}$ & $2.78\cdot10^{-5}$& $ 1.26\cdot10^{-2}$ \\
\hline
2            & 1  & 1.70   & 524 288 & 524 288 & $6.19\cdot10^{-7}$ & $1.77\cdot10^{-4}$& $ 1.01\cdot10^{-1}$ \\
\hline
\end{tabular}
\end{center}
\end{table}

Table~\ref{CPU} shows that, although the order of the error does not show a clear asymptotic behavior, 
there is an obvious decrease of the errors with finer mesh widths. We notice that the maximum of $\sigma$
at $\tau$=210 on the coarse grid is approximately 2. The maximum relative error therefore becomes
negligibly small for mesh spacings of 1/2 or smaller. For coarser grids the error however is significant,
showing that the parameters used here do require a higher accuracy than that of 2 used in~\cite{arr2002}
or 1 used in~\cite{roc2002}.

Notice that  the number of grid points $N_\phi$ used to compute the electric potential
is the same for $\Delta z^f$=1/4 and 1/8, which implies that in fact no grid with mesh spacing
of 1/8 is used in the latter case. Apparently, a mesh width of 1/4 is sufficient for an accurate solution
of the Poisson equation with this choice of the tolerance.

The uniform grid calculations clearly illustrate the gain in efficiency of the refinements. Two orders
of magnitude are gained for the computational memory. The gain of in computation time is less 
accentuated. This is due to the time gained by using less grid points being partially spent
in the refinement procedure. On a regular pc with 1 Gb of memory, the gain in CPU is a factor 
2.5 more pronounced. This is due to limitations in memory, due to which the computer has to swap, which slows
down considerably the computations. Finally, the errors for $\Delta z^f$ are of the same order 
in the uniform and in the refined case, which confirms the good performance of the refinement algorithm. 

A uniform grid with mesh spacing less than 1 could not be obtained for these specific 
simulation parameters. However, the errors with refinements show that finer grids are
necessary for the solution to be reasonably accurate. We can therefore conclude that this refinement procedure
allows us to gain computational time, but, more importantly, to gain so much computational memory that
we are now able to reach a sufficient accuracy.

\section{Summary and conclusions}

We have presented an adaptive grid refinement strategy 
for the computation of negative streamers within the minimal model
in a three dimensional geometry with cylindrical symmetry. 
The equations are rewritten
in dimensionless quantities allowing the transcription 
of the results to arbitrary gases of arbitrary pressure. 

The numerical discretization are based on finite volume methods. It 
uses a second order central scheme for the electron diffusion, and a 
flux limiting scheme for the advection, so that the numerical diffusion is reduced 
and no spurious oscillations are introduced. The Poisson
equation is discretized with a second order central scheme. The time stepping
is achieved with an explicit two-stage Runge-Kutta method.

The explicit time stepping method 
allows us to refine separately the grids for
the continuity equations and those for the Poisson equation. 
The refinement criterion of the continuity equations
is based on a curvature monitor of the solution, while that 
of the Poisson equation is based on an error estimation. 
It results in two series of nested grids, one for the continuity 
and one for the Poisson equation, that communicate with each other 
using adequate restrictions and prolongations of the
densities and the electric field. The refinement method 
has been implemented in such a way that mass, and
therefore charge, is conserved during the refinements. 

Tests on planar ionization fronts show that the leading edge 
has to be included in the refinements to capture the front velocity 
well. This is because the front penetrates a non-ionized high field region
that is linearly unstable against even infinitesimal electron densities;
in fact, the ionization front is a so-called pulled front whose
velocity is determined in the linear leading edge region, 
and not in the nonlinear high gradient region of the front. 
A test on a genuine streamer emerging from a small Gaussian 
ionization seed in a relatively high background electric field 
shows that our refinement including the leading edge region 
performs very well. The space charge layer with its steep 
spatial gradients is also captured well by our refinements. 
Moreover the requested computational memory 
is three orders of magnitude smaller than in a uniform grid 
computation as performed in \cite{roc2002}.

The algorithm enables us to investigate streamers 
with sufficient accuracy in new parameter regimes, namely in 
larger systems and/or in higher electric fields than previously.
The results show how negative streamers increasingly enhance
the field at their tip, both in lower and in higher background fields. 
The spatial gradients increase with the field~\cite{ebe1997} 
and therefore require an increasing spatial accuracy. 

For simulating streamers in a background electric field as high as 0.5 
in dimensionless units, we here used grids with meshes as fine as 1/8, 
much finer than the grids with meshes of 2 and 1 used 
previously~\cite{arr2002,roc2002}. Earlier simulations of negative 
streamers were carried out in a background field of 0.25,
with a grid size of at least 2~\cite{vit1994}. 
These simulations show a field enhancement up
to a value of 1, and our tests show that a finer grid size 
is required for reliable results. 

In fact, since the characteristic scale of the inner structure 
of the streamer front is set by the ionization length 
$\alpha(|\E|_{max})$~\cite{ebe1997,arr2004}, one should take
care that the grid size is fine enough to capture this length scale.
As a rule, we suggest that the grid size should be at least 
four times smaller than the ionization length
in the maximal electric field. For the case of a background field of 0.15,
where the enhanced field grows up to 0.4, this implies a 
grid size not exceed 3 (where we used 2).
For a background field of 0.5, the enhanced field grows up to 2;
the rule then implies that the grid size should not exceed 1/2
(where we actually used 1/4 and 1/8).

Summarizing, we now have an efficient and reliable tool for simulating
negative streamers within the minimal model up to the moment of branching
which will be exploited in future studies of streamer physics. 
Additional effects like electron attachment or photoionization are
important in more complex gases like air, in particular, for positive
streamers. Such effects can be implemented in continuum approximation
along the same lines. 

We finish with an outlook on open problems. First, fully three-dimensional
simulations are required to follow streamer evolution after branching.
We actually expect the branching time in cylindrical symmetry to be a close
upper bound for the actual branching time in the fully three-dimensional
situation~\cite{ebe2002-2,ebe2005}.
Second, numerical solutions of our deterministic fluid model 
show that it is actually admissible for streamer propagation to neglect 
densities below the threshold of 1 particle 
per mm$^3$ where the continuum approximation definitely ceases to hold;
we have used this threshold on our computations with refinement.
In this region which actually belongs to the leading edge of the pulled
ionization front, the discrete nature of particles should be taken into 
account, such work is presently in progress.

%\appendix
\section*{Appendix A: Testing the {\sc Fishpak} fast
Poisson solver}

The {\sc Fishpak} routine is used to solve the Poisson equation. It
is a fast Poisson solver but has limitations with regards to the number of grid points.
We illustrate the instabilities with the {\sc Fishpak} routine on large
grids by an example. Consider the Laplace equation in a
radially symmetric coordinate system $(r,z)\in(0,L_r)\times(0,L_z)$,
\be
\left\{ \begin{array}{l}
\displaystyle
\nabla^2\phi  =  
\Big(-6A + 4A^2 \big(r^2+(z-\half{L_z})^2\big)\Big)
\, e^{-A \left( r^2 + (z-\frac{1}{2}L_z)^2 \right)}  \, , 
\\
\displaystyle
\phi(r,0) =  \phi(r,L_z) = e^{-A \left( r^2 + (z-\frac{1}{2}L_z)^2 \right)}
  \, ,
\\
\displaystyle\frac{\partial\phi}{\partial r}(0,z)  =  0 \, , \quad
\frac{\partial\phi}{\partial r}(L_r,z)  =
-2rA\, e^{-A \left( r^2 + (z-\frac{1}{2}L_z)^2 \right)}
\,,
\end{array} \right.
\label{lapl}
\ee
with $L_r=L_z=1$, giving the analytical solution 
$\phi(r,z)=e^{-A(r^2+(z-L_z/2)^2)}$.

\begin{figure}[t]
\begin{center}
\includegraphics[width=10cm]{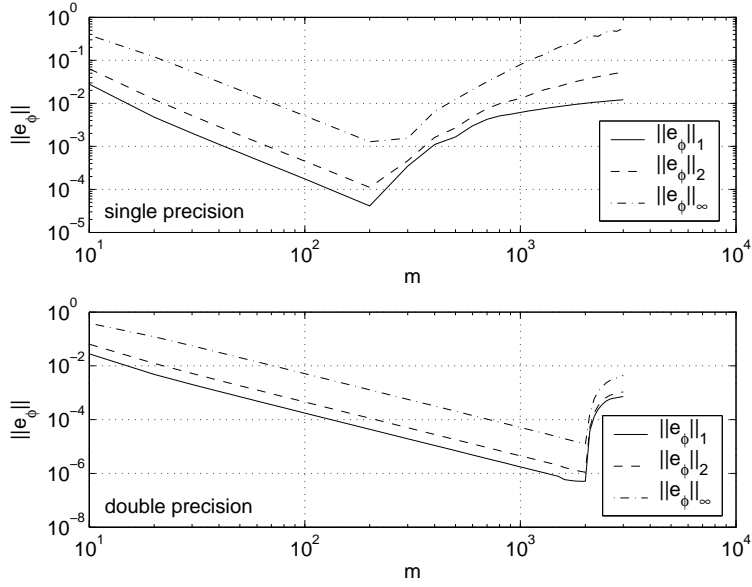}
\caption{\footnotesize\sf The $L_1$-errors (solid), $L_2$-errors (dashed) and 
$L_\infty$-errors (dash-dotted) for $\phi$ in (\ref{lapl}) on 
$m\times m$ grids, as a function of $m$. 
Upper panel, single precision; lower panel, double precision. }
\label{errp}
\end{center}
\end{figure}

\begin{figure}[t]
\begin{center}
\includegraphics[width=10cm]{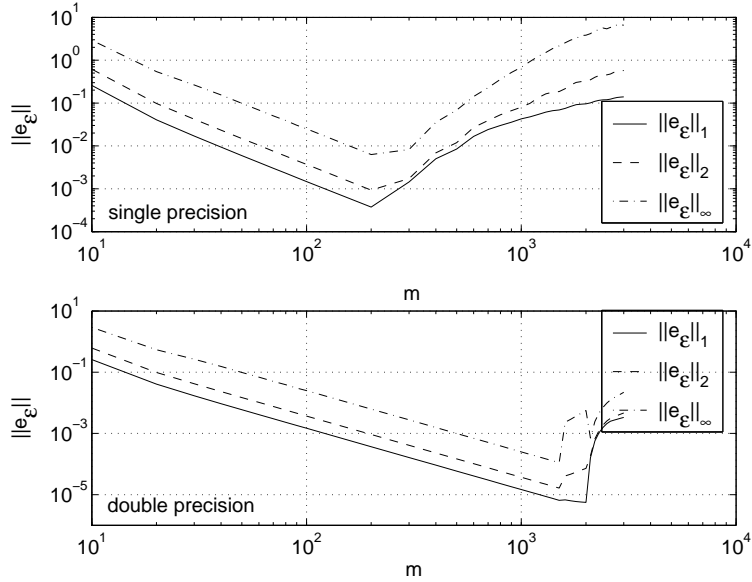}
\caption{\footnotesize\sf The $L_1$-errors (solid), $L_2$-errors (dashed) and $L_\infty$-errors
(dash-dotted) for $|\E|$ in Eq.\,(\ref{lapl}) on $m\times m$ grids, as a function
of $m$. Upper panel, single precision; lower panel, double precision. }
\label{erre} \end{center}
\end{figure}

The accuracy of the method can be characterized by the discrete 
$L_1$, $L_2$ and $L_\infty$-norms of the errors given in Eq.~(\ref{errL}). 
Figure\,\ref{errp} shows these $L_p$-norms of the numerical error $e_{\phi}$
of the results obtained with the {\sc Fishpak} routine on a 
$(m\times m)$-grid, as a function of $m$, with $A=100$.
The upper and lower panel show the error of the single respectively 
double precision computations.
In a similar way we determined the $L_p$-norms of the numerical error 
in the electric field strength $|\tilde{\E}|_{ij}$, computed with 
Eq.\,(\ref{edisc}). They are shown in Figure\,\ref{erre}. 

Up to a number of grid points of approximately 2000, the errors in the electric
potential are of second order, in agreement with
the discretization. The errors in the field are also of second order, even though,
at first sight, a first-order behavior might be expected, since it's the derivative
of a quantity of second order accuracy. This can be explained through an asymptotic
error expansion, which, for simplicity, will be carried in one-dimension.
The second order discretization for $\phi$ will give
\be
\phi_j=\phi(z_j)+\Delta z^2v(z_j)+O(\Delta z^4)\, ,
\ee
with a principal error function $v$ which will be smooth if the solution $\phi$ is so.
By considering local truncation errors it is seen that $v$ should satisfy
$\partial_{zz}v=-\partial_z^4\phi/12$ with corresponding homogeneous boundary conditions.
Then the discretized value of the field at the cell boundary becomes:
\ba
{\mc E}_{j+\frac{1}{2}} = \frac{\phi_j-\phi_{j+1}}{\Delta z}
= \frac{\phi(z_j)-\phi(z_{j+1})}{\Delta z}
+\Delta z(v(z_j)-v(z_{j+1}))+O(\Delta z^3) ,
\ea
which, using a Taylor expansion around $z_{j+\frac{1}{2}}$, yields
\ba
{\mc E}_{j+\frac{1}{2}} = 
-\partial_z\phi(z_{j+\frac{1}{2}}) + O(\Delta z^2)
-\Delta z^2\partial_zv(z_j) + O(\Delta z^3)
= {\mc E}(z_{j+\frac{1}{2}})+O(\Delta z^2)
\ea

It is clear that the performance of the {\sc Fishpak} routine decreases
dramatically when more than approximately 2000 grid points (in double 
precision)are used. We note that these numerical instabilities also show up
on a $m_r\times m_z$ grid if either $m_r$ or $m_z$ are larger than 2000,
approximately, in double precision.

\section*{Appendix B: 
Analytical expressions for the planar front velocity}

The asymptotic velocity (\ref{asvel}) and the convergence towards it
can be derived analytically, following arguments in \cite{ebe2000}.
First, the analysis in \cite{ebe1997} shows that a negative streamer
front is a pulled front whose dynamics is determined in
the leading edge. Linearizing the one-dimensional streamer equations 
in the leading edge around the field $\mE_b<0$ ahead of the front gives
\be
\frac{\partial\sigma}{\partial\tau}=\mE_b\frac{\partial\sigma}{\partial z}
+D\frac{\partial^2\sigma}{\partial z^2}+\sigma f(|\mE_b|)\, ,
\label{lin1d}
\ee
where $f(|\mE_b|)=|\mE_b|\exp(-1/|\mE_b|)$. As the nonlinear front region 
has to be ``pulled along'' by the evolution of the linear perturbations,
the evolution of the linearized equation (\ref{lin1d}) yields an
upper bound for the velocity at each instant of time.

For an initial condition of Gaussian form
\be
\sigma(z,0)=\sigma_0\;\exp\left(-\frac{(z-z_b)^2}{R_b^2}\right),
\ee
as used in the paper, the linearized equation~(\ref{lin1d}) is solved through
\be
\sigma(z,\tau)=\sigma_0\;\exp\Big(f(\mE_b)\tau\Big)\;\;\;
\sqrt{\frac{R_b^2}{R_b^2+4D\tau}}\;
\exp\left(-\frac{(z-z_b+\mE_b\tau)^2}{R_b^2+4D\tau}\right)
\ee
for all times $\tau\ge0$. Solving this equation for $z$, 
the position $z_f(\tau)$ of a fixed electron density level $\sigma_f$ is
\ba
&&\sigma\left(z_f(\tau),\tau\right)=\sigma_f,\\
z_f^\pm(\tau)&=&
z_b+|\mE_b|\tau\pm\sqrt{R_b^2+4D\tau}\,
\sqrt{f(|\mE_b|)\tau-\ln\frac{\sigma_f}{\sigma_0}
      -\frac{1}{2}\ln\frac{R_b^2+4D\tau}{R_b^2}}\\
&\stackrel{\tau\gg1}{\longrightarrow}&
z_b+\left(|\mE_b|\pm\sqrt{4D\;f(|\mE_b|)}\right)\;\tau 
+ O\left(\sqrt{\tau}\right).
\ea
For the pulled front moving to the right, we need $z_f^+(\tau)$. 
The velocity of the level $\sigma_f$ is 
\be
\label{vflin}
v_f^{lin}(\tau)=\partial_\tau z_f^+(\tau).
\ee
This equation determines the asymptotic velocity (\ref{asvel})
that is actually independent of the parameters $\sigma_0$, $R_b$ and $z_b$
of the initial conditions. Moreover the time dependent velocity
$v_f^{lin}(\tau)$ is an upper bound for the actual velocity
$v_f(\tau)$ of the full nonlinear problem. 
The two velocities are compared in Fig.~6.


\begin{thebibliography}{10}
\expandafter\ifx\csname url\endcsname\relax
  \def\url#1{\texttt{#1}}\fi
\expandafter\ifx\csname urlprefix\endcsname\relax\def\urlprefix{URL }\fi

\bibitem{baz2000}
E.~Bazelyan, Y.~Raizer, Lightning Physics and Lightning Protection, Institute
  of Physics Publishing, Bristol, U.K., 2000.

\bibitem{fra1990}
R.~Franz, R.~Nemzek, J.~Winckler, Television image of a large upward electrical
  discharge above a thunderstorm system, Science 249 (1990) 48--51.

\bibitem{sen1995}
D.~Sentman, E.~Wescott, D.~Osborne, M.~Heavner, Preliminary results from the
  {Sprites94} campaign: Red sprites, Geophys. Res. Lett. 22 (1995) 1205--1209.

\bibitem{ger2000}
E.~Gerken, U.~Inan, C.~Barrington-Leigh, Telescopic imaging of sprites,
  Geophys. Res. Lett. 27 (2000) 2637.

\bibitem{pas2002-2}
V.~Pasko, M.~Stanley, J.~Mathews, U.~Inan, T.~Wood, Electrical discharge from a
  thundercloud top to the lower ionosphere, Nature 416 (2002) 152--154.

\bibitem{cle1989}
J.~Clements, A.~Mizuno, W.~Finney, R.~Davis, Combined removal of {SO$_2$},
  {NO$_x$}, and fly ash from simulated flue gas using pulsed streamer corona,
  IEEE Trans. Ind. Appl. 25 (1989) 62--69.

\bibitem{vel2000}
E.~van Veldhuizen, Electrical Discharges for Environmental Purposes:
  Fundamentals and Applications, Nova Science Publishers, 2000.

\bibitem{sat1996}
M.~Sato, T.~Ohgiyama, J.~Clements, Formation of chemical species and their
  effects on microorganisms using a pulsed high voltage discharge in water,
  IEEE Trans. Ind. Appl. 32 (1996) 106--112.

\bibitem{abo2002}
A.~Abou-Ghazala, S.~Katsuki, K.~Schoenbach, F.~Dobbs, K.~Moreira, Bacterial
  decontamination of water by means of pulsed-corona discharges, IEEE Trans.
  Plasma Sci. 30 (2002) 1449--1453.

\bibitem{nai2004}
S.~Nair, A.~Pemen, K.~Yan, E.~van Heesch, K.~Ptasinki, A.~Drinkenburg, A high
  temperature pulsed corona plasma system for tar removal from biomass derived
  fuel gas, J. Electrostatics 61~(2) (2004) 117--127.

\bibitem{boe2005}
J.~Boeuf, Y.~Lagmich, L.~Pitchford, Electrohydrodynamic force and aerodynamic
  flow acceleration in surface dielectric barrier discharges, in: Proc. XXVII
  Int. Conf. Phen. Ion. Gases, Veldhoven, The Netherlands, 2005.

\bibitem{ebe1997}
U.~Ebert, W.~van Saarloos, C.~Caroli, Propagation and structure of planar
  streamer fronts, Phys.~Rev.~E 55 (1997) 1530--1549.

\bibitem{arr2002}
M.~Array{\'a}s, U.~Ebert, W.~Hundsdorfer, Spontaneous branching of
  anode-directed streamers between planar electrodes, Phys.~Rev.~Lett. 88
  (2002) 174502(1--4).

\bibitem{roc2002}
A.~Rocco, U.~Ebert, W.~Hundsdorfer, Branching of negative streamers in free
  flight, Phys.~Rev.~E 66 (2002) 035102(1--4).

\bibitem{liu2004}
N.~Liu, V.~Pasko, Effects of photoionization on propagation and branching of
  positive and negative streamers in sprites, J. Geophys. Res 109 (2004)
  A04301(1--17).

\bibitem{liu2006}
N.~Liu, V.~Pasko, Effects of photoionization on similarity properties of
  streamers at various pressures in air, J. Phys. D: Appl. Phys 39 (2006)
  327--334.

\bibitem{arr2004}
M.~Array{\'a}s, U.~Ebert, Stability of negative ionization fronts:
  Regularization by electric screening?, Phys.~Rev.~E 69 (2004) 036214(1--10).

\bibitem{meu2004}
B.~Meulenbroek, A.~Rocco, U.~Ebert, Streamer branching rationalized by
  conformal mapping techniques, Phys.~Rev.~E 69 (2004) 067402(1--4).

\bibitem{meu2005}
B.~Meulenbroek, U.~Ebert, L.~Sch{\"a}fer, Regularization of moving boundaries
  in a {Laplacian field by a mixed Dirichlet-Neumann} boundary condition --
  exact results, Phys.~Rev.~Lett. 95 (2005) 195004.

\bibitem{ebe2005}
U.~Ebert, C.~Montijn, T.~Briels, W.~Hundsdorfer, B.~Meulenbroek, A.~Rocco,
  E.~van Veldhuizen, The multiscale nature of streamers, submitted to Plasma
  Sources Sci. Technol.

\bibitem{kul2000}
A.~Kulikovsky, The role of the absorption length of photoionizing radiation in
  streamer dynamics in weak fields: a characteristic scale of ionization
  domain, J.~Phys.~D:~Appl.~Phys. 33 (2000) L5--L7.

\bibitem{pan2001}
S.~Pancheshnyi, S.~Starikovskaia, A.~Starikovskii, Role of photoionization
  processes in propagation of cathode-directed streamer,
  J.~Phys.~D:~Appl.~Phys. 34 (2001) 105--115.

\bibitem{kul2002}
A.~Kulikovsky, Comment on ``{S}pontaneous branching of anode-directed streamers
  between planar electrodes'', Phys.~Rev.~Lett. 89 (2002) 229401.

\bibitem{pan2003}
S.~Pancheshnyi, A.~Starikovskii, Two-dimensional numerical modeling of the
  cathode-directed streamer development in a long gap at high voltage,
  J.~Phys.~D:~Appl.~Phys. 36 (2003) 2683--2691.

\bibitem{ebe2000}
U.~Ebert, W.~van Saarloos, Front propagation into unstable states: universal
  algebraic convergence towards uniformly translating pulled fronts, Physica D
  146 (2000) 1--99.

\bibitem{dha1987}
S.~Dhali, P.~Williams, Two-dimensional studies of streamers in gases, J. Appl.
  Phys 62 (1987) 4696--4706.

\bibitem{rai1991}
Y.~Raizer, Gas Discharge Physics, Springer, Berlin, 1991.

\bibitem{vit1994}
P.~Vitello, B.~Penetrante, J.~Bardsley, Simulation of negative-streamer
  dynamics in nitrogen, Phys.~Rev.~E 49 (1994) 5574--5598.

\bibitem{dut1975}
J.~Dutton, A survey of electron swarm data, J. Phys. Chem. Ref. Data 4 (1975)
  664.

\bibitem{vel2002}
E.~van Veldhuizen, P.~Kemps, W.~Rutgers, Streamer branching under influence of
  the power supply, IEEE Trans. Plasma Sci. 30 (2002) 162--163.

\bibitem{bob1998}
Y.~Bobrov, Y.~Yurghelenas, Application of high-resolution schemes in the
  modeling of ionization waves in gas discharges, Comp. Math. and Math. Physics
  38 (1998) 1652--1661.

\bibitem{hun2003}
W.~Hundsdorfer, J.~Verwer, Numerical Solution of Time-Dependent
  Advection-Diffusion-Reaction Equations, Vol.~33 of Series in Comp. Math.,
  Springer, Berlin, 2003.

\bibitem{bar1986}
M.~Barnes, T.~Colter, M.~Elta, Large-signal time-domain modeling of
  low-pressure rf glow discharges, J. Appl. Phys. 61 (1986) 81--89.

\bibitem{blo1996}
J.~Blom, R.~Trompert, J.~Verwer, Algorithm 758: {VLUGR}2: A vectorizable
  adaptive grid solver for {PDE}s in 2{D}, ACM Trans. Math. Softw. 22 (1996)
  302--328.

\bibitem{got2001}
S.~Gottlieb, C.-W. Shu, E.~Tadmor, Strong stability-preserving high-order time
  discretization methods, SIAM Review 43 (2001) 89--112.

\bibitem{wes2001}
P.~Wesseling, Principles of Computational Fluid Dynamics, Vol.~29 of Series in
  Comp. Math., Springer, Berlin, 2001.

\bibitem{schu1976}
U.~Schumann, R.~Sweet, A direct method for the solution of {P}oisson's equation
  with neumann boundary conditions on a staggered grid of arbitrary size, J.
  Comp. Phys 20 (1976) 171--182.

\bibitem{tro1991}
R.~Trompert, J.~Verwer, A static-regridding method for two-dimensional
  parabolic partial differential equations, Appl. Numer. Math. 8 (1991) 65--90.

\bibitem{wac2005}
J.~Wackers, A nested-grid finite-difference {P}oisson solver for concentrated
  source terms, J. Comp. Appl. Math. 180 (2005) 1--12.

\bibitem{gol1996}
G.~Golub, C.~van Loan, Matrix Computations, 3rd Edition, John Hopkins Univ.
  Press, Baltimore, 1996.

\bibitem{bot1997}
E.~Botta, K.~Dekker, Y.~Notay, A.~van~der Ploeg, C.~Vuik, F.~Wubs, P.~de~Zeeuw,
  How fast the {L}aplace equation was solved in 1995, Appl. Num. Math. 24
  (1997) 439--455.

\bibitem{La00}
J.~Lang, Adaptive Multilevel Solution of Nonlinear Parabolic PDE Systems.
  Theory, Algorithm, and Applications, Vol.~16, Springer, 2000.

\bibitem{CDW99b}
P.~Colella, M.~Dorr, D.~Wake, Numerical solution of plasma fluid equations
  using locally refined grids, J. Comp. Phys 152 (1999) 550--583.

\bibitem{CDW99a}
P.~Colella, M.~Dorr, D.~Wake, A conservative finite difference method for the
  numerical solution of plasma fluid equations, J. Comp. Phys 149 (1999)
  168--193.

\bibitem{LPR98}
S.~Li, L.~Petzold, Y.~Ren, Stability of moving mesh systems of partial
  differential equations, SIAM J. Sci. Comp. 20 (1998) 719--738.

\bibitem{QS98}
Y.~Qui, D.~Sloan, Numerical solution of fisher's equation using a moving mesh
  method, J. Comp. Phys 146.

\bibitem{mon2005-1}
C.~Montijn, U.~Ebert, Avalanche to streamer transition in homogeneous fields,
  Tech. Rep. MAS-E0515, CWI,
  {\\http://www.cwi.nl/ftp/CWIreports/MAS/MAS-E0515.pdf, \\
  http://www.arxiv.org/pdf/physics/0508109} (2005).

\bibitem{ebe2002-2}
U.~Ebert, W.~Hundsdorfer, Reply to the comment of {A.A.
  Kulikovsky}~\cite{kul2002} on ``{S}pontaneous branching of anode-directed
  streamers between planar electrodes'', Phys.~Rev.~Lett. 89 (2002) 229402.

\end{thebibliography}
\end{document}